\documentclass[journal=jacsat,manuscript=letter]{achemso}
\usepackage{mathpazo,epsfig,amssymb,amsmath,color,graphicx,rotating,subfigure,booktabs}
\usepackage{gensymb}
\usepackage{colortbl}
\usepackage{graphics}
\usepackage{xcolor}
\usepackage{hyperref}
\usepackage{graphicx,epstopdf}
\usepackage{multirow}
\usepackage{amsmath}
\usepackage{tikz} 
\usepackage{booktabs}
\usepackage[normalem]{ulem}
\usepackage[dvipsnames]{xcolor}
\usepackage{float}
\usepackage[utf8]{inputenc}
\usepackage{amssymb}
\usepackage{amsmath}
\usepackage{ulem}

\usepackage[version=3]{mhchem} 
\usepackage{titlesec}
\usepackage{hyperref}
\title{Recent Developments and Perspectives in Variational Quantum Eigensolvers for Molecular Electronic Structure: Methods, Tradeoffs, and Benchmarking}
 \author{Taylor Harville}
\affiliation
{Chemical Sciences and Engineering Division, Argonne National Laboratory, Lemont, IL 60439, United States}
\author{Rishu Khurana}
 \affiliation{Department of Chemistry, University of Chicago, Chicago, IL 60637, United States}
\alsoaffiliation
{Chemical Sciences and Engineering Division, Argonne National Laboratory, Lemont, IL 60439, United States}
\author{Vitor F. Grizzi}
\affiliation
{Chemical Sciences and Engineering Division, Argonne National Laboratory, Lemont, IL 60439, United States}
\author{Cong Liu}
\email{congliu@anl.gov}
\affiliation
{Chemical Sciences and Engineering Division, Argonne National Laboratory, Lemont, IL 60439, United States}
\alsoaffiliation
{Department of Chemistry, University of Chicago, Chicago, IL 60637, United States}
\alsoaffiliation{Corresponding Author}

\begin{document}
\begin{abstract}
The variational quantum eigensolver (VQE) is a hybrid quantum–classical algorithm designed for noisy intermediate-scale quantum (NISQ) hardware to estimate eigenvalues of many-body Hamiltonians. Unlike fully quantum approaches such as quantum phase estimation (QPE), VQE trades deep coherent circuits for repeated state preparation, measurement, and classical optimization, making it more compatible with limited qubit counts and finite coherence times. Recent developments have focused on reducing quantum resource requirements while retaining chemically meaningful wavefunction structure. In this paper, we examine recent progress in VQE methods for molecular electronic structure with an emphasis on three themes: (i) strategies for circuit and ansatz complexity reduction, including adaptive and selectively screened approaches, (ii) chemically motivated workflows that combine VQE with orbital optimization, fragmentation, and localized active-space ideas to better address strong correlation, and (iii) extensions of VQE to excited-state calculations. Throughout, we emphasize the tradeoffs among parameter count, gate depth, symmetry preservation, measurement overhead, and classical preprocessing, and discuss where these approaches may become most useful for chemically challenging active spaces. We also highlight benchmarking considerations for assessing both accuracy and resource requirements, and conclude with a perspective on regimes in which VQE may offer the greatest long-term value, particularly multireference active spaces and low-lying excited-state manifolds.
\end{abstract}
\section{Introduction}

First-principles, or \textit{ab initio}, calculations have been a staple of computational chemistry for decades and have been central to understanding the properties of molecules and materials. Although \textit{ab initio} electronic structure theory has transformed molecular and materials modeling, the exact solution of the electronic Schr{\"o}dinger equation in a finite orbital basis remains exponentially costly. Recent advances in high-performance parallel computing have expanded the upper limits for molecular sizes, but the growth of classical computing hardware cannot scale fast enough to overcome the costly scaling of quantum chemistry methods.\cite{Feynman1982} This challenge has helped motivate quantum computing as a possible route to simulating many-body quantum systems more efficiently than is possible classically. In quantum chemistry, this idea is especially compelling because the many-body wavefunction grows combinatorially with system size, making strongly correlated electronic structure problems natural targets for quantum algorithms.  

Practical quantum advantage in chemistry will require solving classically intractable problems at chemically meaningful accuracy. However, current quantum hardware remains in the noisy intermediate-scale quantum (NISQ) regime, with limited qubit counts, finite coherence times, and imperfect two-qubit gates. These constraints strongly favor hybrid quantum-classical algorithms that reduce circuit depth and increase noise tolerance.\cite{Preskill2018} Fault-tolerant quantum algorithms remain the long-term target, but their resource requirements are still far beyond current hardware for chemically relevant active spaces.\cite{Elfving2020} 

A leading example is quantum phase estimation (QPE), which in principle offers an exponential speedup for certain strongly correlated problems.\cite{kitaev1995,Abrams1999,Nielsen2010}  In practice, however, QPE requires deep coherent circuits and error correction, placing it well beyond the reach of present-day NISQ devices for molecular applications.\cite{Toyer2017,Aharonov1996,Elfving2020,Dobsicek2007,Wang2019} For broader discussions of QPE theory and resource estimates, we refer the reader to Bauer et al.\cite{Bauer2020}

\begin{figure*}[h!]
\caption{General schematic of the VQE algorithm, showing the fermionic to qubit Hamiltonian mapping and highlighting the interplay between classical and quantum computers. The first step is to map the fermionic Hamiltonian calculated through a Hartree-Fock calculation to a qubit representation of the Hamiltonian. The ansatz is initialized with starting parameters $\overrightarrow{\theta_0}$. The ansatz is then prepared on the quantum computer as a set of gates. This begins the iterative cycle between the quantum and classical computers. Adapted from reference\citenum{Fedorov2022}. Copyright 2022 by Springer Nature under the \href{https://creativecommons.org/licenses/}{Creative Commons CC BY} license.}
\centering
\includegraphics[width=1.0\textwidth]{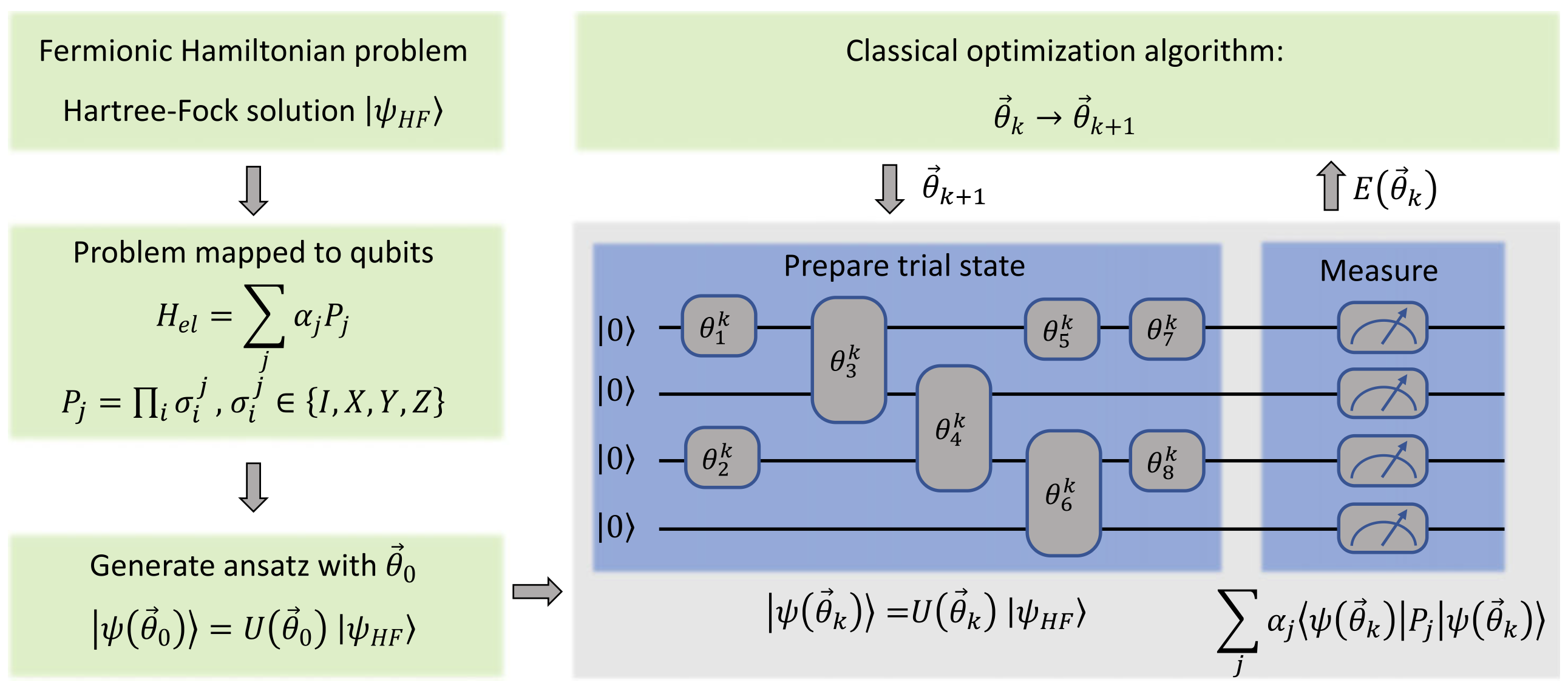}
\label{Vanilla_VQE}
\end{figure*}

Among NISQ-compatible strategies, hybrid quantum-classical methods attempt to trade deep coherent evolution for repeated state preparation, measurement, and classical post-processing. One of the most widely studied examples is the variational quantum eigensolver (VQE), which approximates eigenstates. of a Hamiltonian by minimizing the energy expectation value of a parameterized trial state.\cite{Peruzzo2014} The quantum computer prepares and measures the ansatz state, while the classical optimizer updates the variational parameters to lower the energy according to the variational principle. This hybrid workflow is illustrated schematically in Figure \ref{Vanilla_VQE}. Compared with fully quantum algorithms, this hybrid approach reduces circuit depth at the cost of increased classical optimization and measurement effort. \cite{McClean_2014} The energy is given by
\begin{equation}
    E(\vec{\theta}) = \langle \psi(\vec{\theta}) | \hat{H} | \psi(\vec{\theta}) \rangle
\end{equation}
where $\hat{H}$ is the qubit Hamiltonian, $ | \psi(\vec{\theta}) \rangle $ is the trial wavefunction parameterized by $ \vec{\theta} $, and $ E(\vec{\theta}) $ is the energy expectation value of the trial state. For molecular applications, the electronic Hamiltonian is typically expressed in second quantization
\begin{equation}
    \hat{H} = \sum_{p,q}h_{pq}a^{\dagger}_{p}a_{q} + \frac{1}{2}\sum_{p,q,r,s}g_{pqrs}a^{\dagger}_{p}a^{\dagger}_{q}a_{r}a_{s}
\end{equation}
where $a^{\dagger}_{p}$ and $a_{p}$ are fermionic creation and annihilation operators, respectively, and $h_{pq}$ and $g_{pqrs}$ are one- and two-electron integrals in the chosen orbital basis. 

After fermion-to-qubit mapping, the resulting qubit Hamiltonian can be written as
\begin{equation}
    \hat{H} = \sum_{j}c_{j}\hat{P_{j}}
\end{equation}
where each $\hat{P_{j}}$ is a tensor product of Pauli operators and $c_{j}$ is a real coefficient. Combining the qubit Hamiltonian with the energy expectation value, the quantum computer estimates the energy expectation value by measuring:
\begin{equation}
    E(\vec{\theta}) = \langle \psi(\vec{\theta}) | \hat{H} | \psi(\vec{\theta}) \rangle = \sum_{j}c_j \langle \psi(\vec{\theta}) | \hat{P_{j}}| \psi(\vec{\theta}) \rangle
    \label{vqe_equation_standard}
\end{equation}
A quantum processor estimates the Pauli-string expectation values, and a classical optimizer updates $\vec{\theta}$ to minimize the energy.

VQE's flexibility stems from the choice of ansatz, which influences both the quantum circuit complexity and the classical optimizations. Broadly, VQE ansatzes can be divided into chemistry-inspired forms, which preserve more of the structure of electronic wavefunction theory, and hardware-efficient forms, which prioritize shallow implementation on specific devices. For quantum chemistry applications, chemistry-inspired ansatzes are often preferred because they more naturally preserve physically relevant structure such as particle number, spin, and excitation hierarchy. Accordingly, this review focuses primarily on chemistry-inspired VQE approaches and their recent extensions. 

The standard chemistry-inspired baseline is the unitary coupled-cluster ansatz with singles and doubles (UCCSD), usually built from a Hartree–Fock reference.\cite{Peruzzo2014}
\begin{equation}
    | \psi(\vec{\theta}) \rangle = e^{\hat{T}(\vec{\theta})-\hat{T^{\dagger}}(\vec{\theta})} |\phi_{ref} \rangle
\end{equation}
where the cluster operator, $\hat{T}$, is the sum of excitation operators with excitation levels indexed by $i$
\begin{equation}
    \hat{T}=\sum_i \hat{T_i }
\end{equation}
In UCCSD, this expansion is truncated at single and double excitations,
\begin{equation}
    \hat{T}_{UCCSD}=\hat{T}_1+\hat{T}_2
\end{equation}
with
\begin{equation}
    \hat{T}_{1}= \sum t_i^a a^{\dagger}_{a}a_{i} \text{  ,  } \hat{T}_2 = \frac{1}{4}\sum t_{ij}^{ab} a^{\dagger}_{a}a^{\dagger}_{b}a_{i}a_{j}
\end{equation}
Here, the indices i,j,k,... and a,b,c,.. denote occupied and virtual molecular orbitals respectively, and $t_i^a$ and $t_{ij}^{ab}$ are the variational cluster amplitudes. Although UCCSD provides a physically meaningful starting point, its implementation can become costly as system size grows, particularly when measured in terms of two-qubit gate count, parameter number, and measurement overhead. These limitations have motivated the development of more compact, adaptive, and chemically structured VQE variants.

In our view, the most compelling chemical targets for VQE are not routine single-reference equilibrium molecules, where classical methods remain more efficient, but rather multireference active spaces and low-lying excited-state manifolds, where classical treatments become more delicate, expensive, or strongly method-dependent. 
Future targets for VQE to demonstrate quantum advantage include catalytic systems requiring large active spaces and open-shell transition-metal complexes. In these regimes, classical methods can encounter bottlenecks or memory limitations that prohibit detailed analysis.\cite{Levine2020,Montgomery2018,Markus2017} Hybrid algorithms, such as VQE, which encode electron correlation more naturally on quantum hardware, may scale more favorably with active-space size and could be used to study these previously intractable systems. Although chemically useful quantum advantage is unlikely in the immediate term, continued development of VQE algorithms remains important because future progress will depend not only on hardware improvements, but also on software strategies that reduce measurements, preserve symmetry, and focus quantum resources on the most chemically complex parts of the problem.\cite{Bal2024,Jiang2025,Awshalom2025}

Earlier surveys of variational quantum algorithms, most notably the extensive treatment by Tilly et al.\cite{Tilly2022}, cataloged the foundational landscape of VQE up to early 2022. Rather than attempting an exhaustive catalog of all recent VQE variants, we focus on developments most relevant to molecular electronic structure and organize them by the computational bottlenecks they target. Specifically, we organize the discussion around three themes: (i) circuit complexity reduction methods, including adaptive and selectively screened approaches, (ii) chemically motivated workflows that combine VQE with orbital optimization, fragmentation, and localized active-space ideas to better address strong correlation, and (iii) extensions of VQE to excited-state calculations. Across all three themes, we emphasize that apparent savings in one resource—such as parameter count or circuit depth—often come at the cost of increased measurements, more classical preprocessing, or stronger chemical approximations. 

This paper is organized as follows. Section \ref{ansatz_compression} discusses ansatz compression and circuit complexity reduction. Section \ref{chemicall_motivated} examines chemically motivated VQE workflows. Section \ref{excited_state_methods} reviews excited-state extensions of VQE. Section \ref{benchmarking} presents illustrative benchmarks comparing representative ansatz-construction strategies. Section \ref{conclusions} summarizes the main conclusions, and Section \ref{perspective} offers a perspective on where VQE may provide the greatest future value in chemistry.

\section{Ansatz Compression and Circuit-complexity Reduction Methods}
\label{ansatz_compression}
%\newline
The methods in this section aim to reduce the practical cost of VQE on near-term hardware by compressing the ansatz, lowering gate depth, or reducing the number of entangling operations. In molecular applications, however, these savings are rarely one-dimensional. A method that lowers the final circuit depth may require repeated gradient measurements, more intensive classical preprocessing, or less direct control over physical symmetries. We therefore organize this section not as a list of isolated algorithms, but by the main strategy used to reduce quantum cost: adaptive ansatz growth, classical prescreening, and improved expressivity at fixed circuit depth.

\subsection{Adaptive ansatz growth: ADAPT-VQE and qubit-ADAPT-VQE}

A major limitation of fixed ansatzes such as UCCSD is that they include many operators whose contributions to the correlated wavefunction may be insignificant. Adaptive ansatz methods address this by constructing the wavefunction iteratively, adding only those operators that are considered most important for the current state. The best known example is the Adaptive Derivative-Assembled Pseudo-Trotter Variational Quantum Eigensolver (ADAPT-VQE), which dynamically builds the ansatz from a predefined operator pool by repeatedly selecting the operator with the largest energy-gradient contribution.\cite{Grimsley2019,Preskill2018}

\begin{figure*}[h!]
\caption{The ADAPT-VQE algorithm begins on the quantum computer by generating a pool of the possible qubit operators. The computed VQE ansatz is grown by one operator per iteration. The operator with the largest contribution to the correlation energy (determined by magnitude of the measured gradient) is added each iteration. This procedure is repeated, and the ansatz grown, until the convergence tolerance is satisfied. This algorithm neglects terms with near zero contributions to the energy, reducing ansatz size without sacrificing accuracy. Adapted from reference \citenum{Grimsley2019}. Copyright 2019 by Nature Communication under the \href{https://creativecommons.org/licenses/}{Creative Commons CC BY} license.}
\centering
\includegraphics[width=1.0\textwidth]{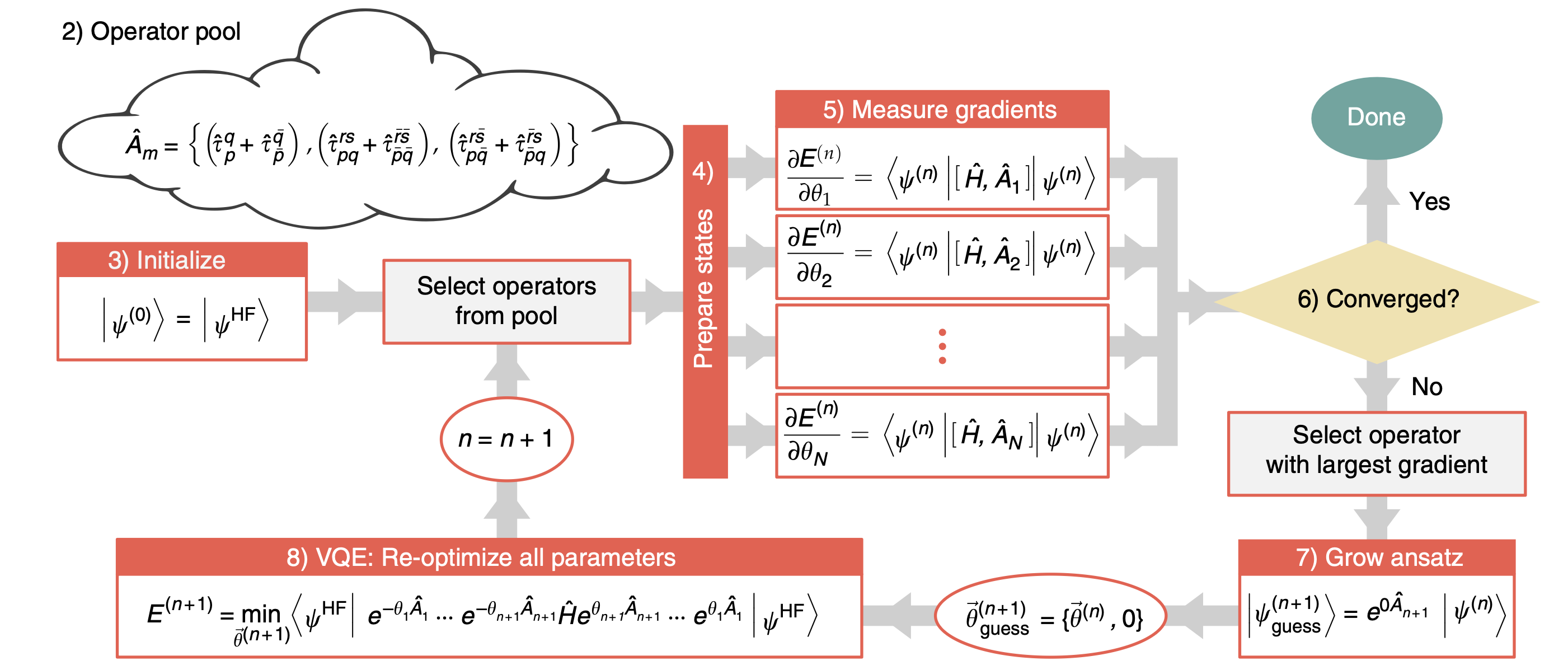}
\label{ADAPT_VQE}
\end{figure*}

At each iteration, ADAPT-VQE ranks candidate operators according to the magnitude of their energy gradients with respect to the current ansatz state, commonly evaluated through commutators with the Hamiltonian,
    \begin{equation}
       \frac{\partial E}{\partial \theta_k} = \langle \psi | [ \hat{H}, \hat{A}   _k ] | \psi \rangle, \textbf{}
    \end{equation}
where $ \hat{H} $ is the molecular Hamiltonian, $\hat{A}_k$ is k-th operator from the pool, and $\theta_k$ is its associated variational parameter. For each iteration, the operator with the largest gradient magnitude is added to the ansatz and all current parameters are then re-optimized in a full VQE step. This process repeats until the norm of the gradient, $\|\vec{g}\| $, drops below a predefined threshold indicating that the remaining operators are expected to contribute only weakly to the energy as shown in Figure \ref{ADAPT_VQE}. \cite{Grimsley2019}

    \begin{equation}
         \|\vec{g}\| = \sqrt{\sum_i \left(\frac{\partial E}{\partial \theta_i} \right)^2}
    \end{equation}

A major advantage of ADAPT-VQE is its ability to construct compact ansatzes with fewer variational parameters and quantum gates than standard UCCSD-VQE. In its fermionic form, ADAPT-VQE typically uses a pool of generalized excitation operators. Unlike conventional UCCSD, this pool is not restricted to a fixed occupied-to-virtual hierarchy defined by the Hartree–Fock reference, and can therefore identify more compact operator sequences that better adapt to changes in correlation along a potential-energy surface.  

For small benchmark systems such as LiH and BeH$_2$, ADAPT-VQE has been shown to achieve chemical accuracy (typically taken as approximately 1 kcal/mol or 1.6 mHa, relative to full configuration interaction (FCI)) with significantly fewer operators than UCCSD-VQE, and often maintains better accuracy away from equilibrium geometries where fixed reference-based ansatzes become less balanced. These compact ansatzes can reduce both parameter count and final circuit depth relative to UCCSD.\cite{Grimsley2019} 

ADAPT-VQE has inspired various iterative quantum computing algorithms.\cite{Tang2021,Ramoa2025,Fitzpatrick_2024} A closely related variant is qubit-ADAPT-VQE, which retains the same adaptive logic but replaces the fermionic operator pool with a pool of Pauli-string generators defined directly in qubit space. The motivation is hardware-oriented: by selecting operators after fermion-to-qubit mapping, qubit-ADAPT-VQE can reduce the number of multi-qubit entangling gates needed to implement each selected term. In favorable cases, this leads to shallower circuits and substantially lower CNOT counts than either fermionic ADAPT-VQE or UCCSD-VQE.

The distinction between the two forms of ADAPT-VQE is not simply algorithmic but conceptual. Fermionic ADAPT-VQE more naturally preserves chemically meaningful operator structure and symmetry constraints, whereas qubit-ADAPT-VQE is often more attractive when entangling-gate fidelity is the dominant hardware bottleneck. The price of the qubit-space formulation is that symmetry preservation is no longer automatic unless built into the operator pool, and the number of variational parameters can increase even when the per-operator circuit cost decreases.\cite{Tang2021,Preskill2018,Yordanov2021}

The main tradeoff in both ADAPT variants is that ansatz compactness is achieved through repeated operator screening. Each iteration requires evaluating gradients over the full operator pool, so the cost of constructing the ansatz can become dominant in measurement effort even when the final circuit is short. This measurement burden grows with pool size and can offset the savings from reduced parameter count, especially in larger active spaces or when tight convergence thresholds require many iterations.

Taken together, adaptive ansatz methods illustrate a recurring theme in VQE design: compact final wavefunctions do not come for free. ADAPT-VQE and qubit-ADAPT-VQE often outperform fixed UCCSD-like constructions in ansatz efficiency and potential-energy-surface robustness, but they do so by shifting cost toward repeated gradient measurements and, in the qubit-space case, more careful handling of symmetry and optimization complexity.

\subsection{Improving expressivity without deeper circuits: nuVQE}

A different strategy for reducing effective quantum cost is to increase wavefunction flexibility without lengthening the quantum circuit itself. The non-unitary variational quantum eigensolver (nuVQE) does this by augmenting a standard UCCSD-VQE with a classically parameterized non-unitary Jastrow operator inspired by classical quantum Monte Carlo (QMC) ideas.\cite{Benfenati2021} The resulting hybrid ansatz can improve the effective expressivity of the trial wavefunction without increasing the depth of the underlying quantum circuit. 

In nuVQE, the trial wavefunction is expressed as a non-normalized ansatz resulting from the application of a non-unitary operator $ \hat{O}(\vec{\lambda}) $ followed by a unitary operator $ \hat{U}(\vec{\theta}) $ on an initial state $ |\psi_0\rangle $:
    \begin{equation}
        |\psi_O(\vec{\theta})\rangle = \hat{O}(\vec{\lambda}) \hat{U}(\vec{\theta}) |\psi_0\rangle
    \end{equation}
    where $ |\psi_0\rangle $ is typically the Hartree-Fock state, and $ \vec{\theta} $ and $ \vec{\lambda} $ are variational parameters for the unitary and non-unitary operators, respectively. The energy is computed by:
    \begin{equation}
       E = \frac{\langle \psi_O(\vec{\theta}) | \hat{H} | \psi_O(\vec{\theta}) \rangle}{\langle \psi_O(\vec{\theta}) | \psi_O(\vec{\theta}) \rangle}
    \end{equation}
    which can be explicitly expressed as:
    \begin{equation}
        E = \frac{\langle \psi(\vec{\theta}) | \hat{O}^\dagger(\vec{\lambda}) \hat{H} \hat{O}(\vec{\lambda}) | \psi(\vec{\theta}) \rangle}{\langle \psi(\vec{\theta}) | \hat{O}^\dagger(\vec{\lambda}) \hat{O}(\vec{\lambda}) | \psi(\vec{\theta}) \rangle}
    \end{equation}
    where $ |\psi(\vec{\theta})\rangle = \hat{U}(\vec{\theta}) |\psi_0\rangle $.  The additional Jastrow parameters are optimized classically together with the unitary circuit parameters.

The linearized form of the qubit-level Jastrow factor is

    \begin{equation}
      \hat{J}(\vec{\alpha}, \vec{\lambda}) = 1 - \sum_{i=1}^N \alpha_i Z_i - \sum_{i<j=1}^N \lambda_{i,j} Z_i Z_j
    \end{equation} 
    where the variational parameters $  \alpha_i  $ and $  \lambda_{ij}  $ are optimized classically together with the unitary parameters. In practice, these additional classical parameters can improve energy estimates and may partly compensate for some noise-induced errors on NISQ hardware, but they do not eliminate the need to evaluate the molecular Hamiltonian through Pauli measurements. In other words, nuVQE can improve accuracy at fixed circuit depth, but it does not fundamentally solve the measurement bottleneck.

    Reported studies suggest that nuVQE can substantially improve ground-state energies relative to standard VQE at comparable circuit depth and can provide a built-in error-mitigation effect that partially compensates for decoherence, but the tradeoff is shifted toward classical normalization overhead and, in some implementations, increased statistical uncertainty as the number of effective measured quantities grows.\cite{Tilly2022} This idea has also been extended into localized active-space workflows,\cite{wang2025} discussed further in Section 3, where non-unitary dressing is used to recover part of the inter-fragment correlation missing from mean-field fragment coupling.
    
     \subsection{Classical prescreening and selective ansatz construction: USCC}
 
A second route to ansatz compression is to avoid repeated quantum screening altogether by selecting operators classically before the VQE optimization. Unitary selective coupled cluster (USCC) follows this logic by prescreening operators on the classical computer, in a way inspired by selected configuration interaction (SCI) and heat-bath ideas, retaining only those excitations expected to contribute significantly to the correlated wavefunction.\cite{Holmes2016} 

In contrast to ADAPT-VQE, which ranks operators by measured energy gradients on the quantum device, USCC shifts the screening step to classical preprocessing. It begins from excitations generated with respect to a reference determinant and selectively retains terms above a chosen threshold, with the option to include disconnected higher-order excitations when they are expected to recover missing correlation.

USCC can therefore recover part of the benefit of adaptive ansatz construction  without incurring repeated quantum gradient evaluations. In systems where standard UCCSD-VQE fails to achieve chemical accuracy, USCC has been shown to recover the missing correlation energy through a more selective operator set and, when necessary, through inclusion of the disconnected triple and quadruple excitations that are missing from standard UCCSD. 
 
The tradeoff is that this reduction in measurement overhead does not necessarily lead to shallower final circuits. Because USCC may rely on the higher-order disconnected excitations to recover missing correlation, the selective ansatz can still become expensive to implement in terms of two-qubit gates such as CNOTs. In this sense, USCC exchanges one kind of adaptivity cost for another: less quantum screening, but potentially more complicated circuit blocks. Its practical niche is therefore measurement-limited settings where classical prescreening is inexpensive and final gate count is not the sole bottleneck.
    
\subsection{Other VQE variants}
The rapid pace of VQE development has produced many additional variants aimed at improving optimization robustness or reducing specific sources of overhead. The methods that are most representative of chemistry-relevant directions were discussed in more detail. Other example variants include overlap-based ADAPT variants\cite{Feniou_2023} that modify operator selection to reduce local minima, gradient-free adaptive schemes (e.g., GGA-VQE\cite{Feniou2025}) designed to improve noise resilience, amplitude-reordering approaches (e.g., AR-ADAPT-VQE\cite{lan2022amplitude}) that accelerate ansatz growth, and optimization improvements based on quantum natural gradients\cite{shi2025efficienthamiltonianawarequantumnatural} or machine-learning-guided search.\cite{zhang2025diffusionenhancedoptimizationvariationalquantum} We do not discuss these methods in detail here, but note that they generally fit within the same design space described above: they attempt to reduce one practical bottleneck in VQE while leaving others largely unchanged.

The methods above illustrate a central point that recurs throughout this review: reducing one nominal quantum resource does not automatically reduce the total cost of a VQE calculation. Adaptive methods such as ADAPT-VQE and qubit-ADAPT-VQE reduce ansatz size and often improve potential-energy-surface accuracy relative to fixed UCCSD, but they pay for this compactness through repeated full-pool gradient measurements. nuVQE improves effective wavefunction flexibility without lengthening the circuit, but still requires evaluation of the Hamiltonian and introduces normalization overhead through the non-unitary dressing. USCC shifts operator selection to classical preprocessing and avoids repeated quantum screening, but can require higher-level operators whose circuit implementations are more expensive. 

The practical choice among these methods therefore depends on which bottleneck is dominant. If entangling-gate fidelity is the limiting factor, qubit-space adaptive constructions may be attractive. If measurement cost dominates, classical prescreening may be preferable. If fixed-depth circuits are required but accuracy is insufficient, non-unitary dressing can be useful. For molecular applications, the most informative comparison is not simply parameter count or circuit depth alone, but how a method redistributes cost across measurements, classical preprocessing, symmetry control, and final gate complexity.

 \section{Chemically motivated VQE workflows}
 \label{chemicall_motivated}
%\newline
Beyond ansatz compression alone, several recent developments embed the variational solver within workflows already familiar in electronic structure theory. These approaches are motivated less by hardware efficiency alone than by chemical structure: the choice of orbital basis, the locality of correlation, and the partitioning of large systems into active or weakly coupled subspaces. In practice, they often trade a more elaborate classical workflow for reduced quantum cost and a more chemically meaningful treatment of strong correlation. We group these methods here by the kind of structure they exploit: orbital optimization, fragmentation and localized active spaces, and entanglement-based partitioning. 

\subsection{Orbital optimization around VQE: ADAPT-VQE-SCF}
     
The ADAPT-VQE Self-Consistent Field (ADAPT-VQE-SCF) method extends adaptive VQE into the orbital-optimization framework of the classical complete active-space SCF (CASSCF) method to capture non-dynamical correlation on NISQ devices.\cite{Fitzpatrick_2024} Building on ADAPT-VQE, which iteratively grows the ansatz by selecting operators based on energy gradients, ADAPT-VQE-SCF adds a self-consistent-field (SCF) loop that optimizes molecular orbitals (MOs) via orbital rotations.

\begin{figure*}[h!]
\caption{Schematic diagram of the ADAPT-VQE-SCF algorithm, highlighting the interplay between the solution of the wavefunction via VQE and the orbital optimization of CASSCF.  Reproduced with permission from reference \citenum{Fitzpatrick_2024}. Copyright 2024 by American Chemical Society.}
\centering
\includegraphics[width=1.0\textwidth]{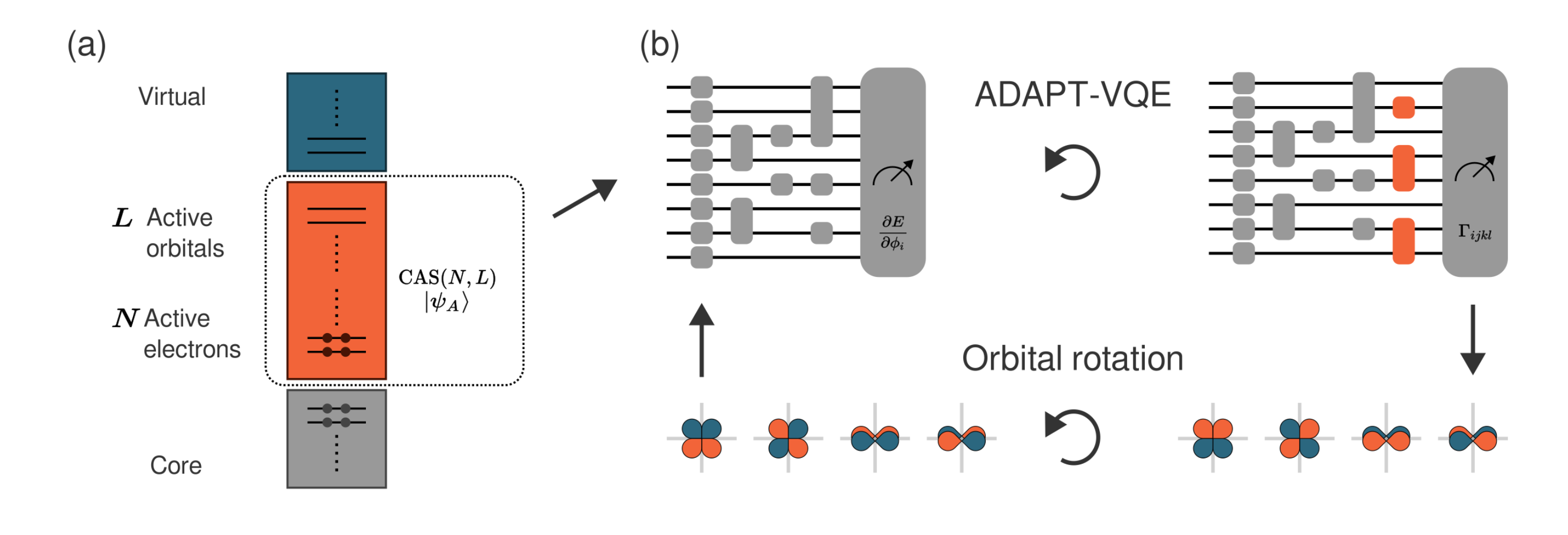}
\label{ADAPT_VQE_SCF}
\end{figure*}

This method is conceptually analogous to CASSCF, which simultaneously optimizes molecular orbitals and configuration interaction coefficients to capture the non-dynamical electron correlation in ground and excited states.\cite{Roos1980,Siegbahn1981} ADAPT-VQE-SCF extends this idea by alternating between minimization of the active-space energy with VQE and the orbital optimization, as illustrated in Figure \ref{ADAPT_VQE_SCF}. This way, the computationally demanding active-space diagonalization is replaced by a quantum variational solver while orbital relaxation is retained on the classical side. To avoid the prohibitive cost of fully converging an ADAPT-VQE wavefunction at every orbital step, the algorithm alternates between single-operator ansatz growth and orbital relaxation. This alternating strategy reduces the total quantum effort while still allowing the orbital basis to adapt to multireference character.

ADAPT-VQE-SCF’s key advantage is its improved treatment of non-dynamical correlation. The commonly used UCCSD ansatz captures dynamical correlation, but cannot account for the multireference character. ADAPT-VQE-SCF can capture this multireference character that is important for studying many transition metal containing catalytic complexes. In the reported ferrocene [Fe(C$_5$H$_5$)$_2$] benchmark with the cc-pVDZ basis (requiring 20 qubits), it achieves chemical accuracy with fewer parameters and shallower circuits  compared to UCCSD.\cite{Fitzpatrick_2024} ADAPT-VQE-SCF’s macro-iteration loop between adaptive VQE and classical orbital optimization incurs an additional classical cost of O(N$^3$) per cycle for the orbital-rotation matrix diagonalization, yet it dramatically improves non-dynamical correlation recovery without increasing quantum resources. The measurement overhead per VQE sub-iteration remains identical to standard ADAPT-VQE (gradient evaluations over a symmetry-adapted pool), but as highlighted in Section 2.1 these recurrent gradient scans across the full operator pool increase the measurement cost and can dominate the wall-clock runtime. This orbital optimized adaptive framework is particularly relevant for transition-metal and open-shell systems, where orbital relaxation can substantially improve the compactness and quality of the variational ansatz, provided that the active-space size permits the extra measurement budget. One limitation of ADAPT-VQE-SCF is the incomplete treatment of dynamical correlation, which is critical for predicting chemical reactions (e.g., reaction energies and barriers). To capture both dynamical and non-dynamical correlation effectively, more sophisticated hybrid multireference strategies will likely be required.

\subsection{Fragmentation and localized active spaces: FMO-VQE and LAS-nuVQE}
     
The Fragment Molecular Orbital-Based Variational Quantum Eigensolver (FMO-VQE) combines VQE with the fragment molecular orbital (FMO) framework to extend quantum simulations to larger molecular systems.\cite{Lim2024} Rather than solving the full Hamiltonian at once, the system is partitioned into embedded monomer and dimer subsystems, each treated with a smaller VQE calculation with costs scaling by fragment size. The total energy is then assembled from fragment contributions in the usual FMO manner.\cite{Barca2020} 

In FMO-VQE, the molecular system is fragmented into sets of monomer and dimers, and the total energy is approximated as:
    \begin{equation}
       E_{\text{FMO}} = \sum_I E_I + \sum_{I<J} \Delta E_{IJ}
    \end{equation}
    where $ E_I $ is the energy of monomer $ I $ computed via VQE in the presence of the electrostatic potential from the rest of the system, and $ \Delta E_{IJ} $ is the dimer correlation energy.
    \begin{equation}
       \Delta E_{IJ} = E_{IJ} - E_I - E_J
    \end{equation}
where $E_{IJ}$ is the energy of fragment dimer composed of monomers $I$ and $J$ computed via VQE in the presence of the electrostatic potential from the rest of the system. Bond breaking across fragments is handled by either Hybrid Orbital Projection (HOP)\cite{Barca2020} or Adapted Frozen Orbital (AFO).\cite{Nakano2000}

The appeal of FMO-VQE is straightforward: fragmentation can reduce qubit requirements dramatically by replacing one large calculation with many smaller embedded ones. Reported benchmarks on hydrogen-chain systems show substantial qubit savings while retaining near-chemical accuracy when correlation is sufficiently local. As in classical FMO, however, accuracy depends strongly on fragment choice, embedding quality, and the treatment of covalent cuts.\cite{Lim2024} For delocalized systems or cases with significant inter-fragment correlation, higher-order fragment corrections or carefully designed bond-breaking schemes may be required.\cite{Federov2008}

The Localized Active-Space Non-Unitary Variational Quantum Eigensolver (LAS-nuVQE) combines localized active-space self-consistent field (LASSCF) ideas with non-unitary variational dressing to target strongly correlated systems with local multireference character. In this approach, the active-space is partitioned into localized fragment subspaces treated independently at the multiconfigurational level, and a non-unitary dressing is then used to recover part of the inter-fragment correlation missing from the mean-field coupling.\cite{wang2025} 

However, LASSCF scales with fragment size, limiting calculations to moderate active spaces per fragment, and approximates long-range correlation via the mean-field approximation, which is potentially inaccurate for delocalized systems. LAS-nuVQE addresses this limitation by applying a non-unitary variational dressing (nuVQE discussed in Section 2.3) on top of the localized active-space description, thereby recovering part of the correlation missing from the mean-field inter-fragment coupling. 

LAS-nuVQE was reported to effectively recover this inter-fragment correlation with shallow circuits (70 gates), achieving chemical accuracy in H$_4$ and cyclobutadiene (C$_4$H$_4$) using 4–8 qubits. It is also shown that LAS-nuVQE outperforms classical LASSCF when compared to FCI, but LAS-nuVQE faces challenges.\cite{wang2025}  Localization approximations may miss long-range correlations in delocalized systems. Compared to ADAPT-VQE-SCF, which optimizes orbitals adaptively but requires more measurements, LAS-nuVQE shifts complexity to classical LASSCF preprocessing. Unlike ADAPT-VQE-SCF, LAS-nuVQE does not iteratively alternate between capturing the inter-fragment correlation and orbital rotations. Just like fully classical CASSCF/LASSCF, while LAS-nuVQE does capture important non-dynamical correlation, it does not fully capture the dynamical correlation often needed to study reaction pathways and catalytic processes. Perturbative corrections such as second-order perturbation theory (PT2) are needed to recover this dynamical correlation. The per-fragment ansatz depth and CNOT count are lower than a fully delocalized ADAPT-VQE treatment due to the fragmentation of the system. For large catalytic systems this approach may offer a practical route to chemical accuracy with manageable quantum resource requirements, with the cost being the addition of the approximation of the inter-fragment correlation that must be validated against full-system benchmarks. As with other fragmentation-based approaches, LAS-nuVQE's performance depends on the physical validity of the chosen fragmentation and may degrade when long-range correlation or strong inter-fragment delocalization becomes important.

Together, FMO-VQE and LAS-nuVQE represent two chemically motivated decomposition strategies with different levels of approximation. FMO-VQE fragments the full molecular problem into embedded subsystem energies, making it attractive for large molecules when the main challenge is system size. LAS-nuVQE instead targets large strongly correlated active spaces by localizing the multireference problem itself, making it more appealing when the main challenge is non-dynamical correlation. In both cases, quantum savings arise by assuming that chemically important correlation is local enough to partition.
\begin{table*}[!htb] 
\caption{Qualitative comparison of the VQE variants discussed in Sections \ref{ansatz_compression} and \ref{chemicall_motivated}. The table emphasizes the central tradeoffs among ansatz compactness, circuit depth, measurement overhead, and chemical applicability.} \label{tab:vqe_compact} 
\centering \scriptsize \setlength{\tabcolsep}{4pt} \renewcommand{\arraystretch}{1.1} \begin{tabular}{|p{1.8cm}| p{2.8cm}| p{4.0cm} |p{4.0cm}| p{2.9cm}|} 
\hline 
\textbf{Method} & \textbf{Type} & \textbf{Main benefit} & \textbf{Main cost} & \textbf{Typical niche} \\ \hline 
ADAPT-VQE & Adaptive fermionic ansatz & Compact ansatz; fewer parameters; improved PES accuracy vs UCCSD & Repeated full-pool gradient measurements & High accuracy in small/moderate systems \\ \hline
qubit-ADAPT-VQE & Adaptive qubit-space ansatz & Lower CNOT count; shallower circuits & More parameters; harder optimization; loss of symmetry control & Hardware limited by entangling-gate errors \\ \hline
nuVQE & Unitary ansatz + Jastrow dressing & Better accuracy at fixed depth; possible error mitigation & Full measurement burden remains; added classical overhead & Shallow-circuit NISQ chemistry \\ \hline
USCC & Classical prescreen adaptive method & Avoids repeated quantum gradient evaluations; higher-order terms  & Higher-order terms can increase gate count & Cases where measurement cost dominates \\ \hline
ADAPT-VQE-SCF & Adaptive VQE + orbital optimization & Better non-dynamical correlation; orbital relaxation & ADAPT measurement cost + SCF iterations & Multi-reference and transition-metal systems \\ \hline
FMO-VQE & Fragment molecular orbital + VQE & Large qubit reduction via fragmentation & Fragment/embedding errors; difficult covalent cuts & Large molecules with local correlation \\ \hline
LAS-nuVQE & Localized active spaces + non-unitary dressing & Low-depth fragment circuits; recovers some inter-fragment correlation & Fragmentation error; incomplete long-range/dynamical correlation & Large locally correlated systems \\\hline 
ClusterVQE & Qubit clustering + dressed Hamiltonian & Smaller per-cluster circuits & Clustering/dressing overhead; may miss long-range entanglement & Systems with localized entanglement \\ \hline 
\end{tabular} 
\end{table*}
\subsection{Entanglement-based partitioning: ClusterVQE}
Not all useful partitions are chemically predefined. ClusterVQE instead decomposes the problem in qubit space, using estimated correlation structure to identify clusters of strongly coupled degrees of freedom. In this sense, it occupies an intermediate position between hardware-motivated partitioning and chemically motivated fragmentation.\cite{Zhang2022}

In ClusterVQE, the qubit space is divided into clusters where each cluster is assigned to a separate quantum circuit. The clusters are determined by maximizing intra-cluster mutual information (MI) in a fashion similar to that of the classical density matrix renormalization group (DMRG) method.\cite{Rissler_2006} The initial reduced density matrix for the system is obtained from the classical Hartree-Fock calculation. The clustering algorithm uses a classical graph partitioning algorithm to minimize the sum of inter-cluster mutual information. Inter-cluster correlations are incorporated by dressing the Hamiltonian, $ H_d $, with an entangler $\hat{U}_{c_{ij}}$ that connects cluster $i$ with cluster $j$.\cite{Zhang2022} 

    \begin{equation}
       \hat{H}_d = \Pi_{i\neq j}\hat{U}^{\dagger}_{ij}\hat{H}\hat{U}_{ij}
    \end{equation}

Residual inter-cluster effects are incorporated through a dressed Hamiltonian, so that each cluster is solved in an effective environment rather than in isolation. This reduces the size of the individual quantum problems, but comes at the cost of increased classical resources required to update the dressed Hamiltonian.

ClusterVQE is intended to enable larger effective problem sizes by reducing per-cluster quantum hardware demands. For LiH (8 qubits, STO-3G basis), it achieves chemical accuracy with clusters of 4 qubits each, using circuits 2–3 times shallower than qubit-ADAPT-VQE and requiring fewer iterations than iQCC.\cite{iQCC} As with ADAPT-VQE (both fermionic and qubit), ClusterVQE iteratively grows the ansatz by comparing contributions of each operator to the correlation energy using analytic gradients.  ClusterVQE can be limited by the classical optimization of clustering and the iterative dressing which introduce additional computational overhead, scaling with system size and cluster count. For strongly correlated systems, incomplete capture of inter-cluster entanglement may lead to convergence issues, requiring more iterations. To date, ClusterVQE has primarily been validated in simulation, so its robustness on real noisy hardware remains unclear.  Additionally, the barren-plateau problem remains a risk if the initial Hartree–Fock reduced density matrix is poorly conditioned. ClusterVQE is most attractive when entanglement is sufficiently localized, but its performance depends strongly on the quality of the clustering and on how completely inter-cluster correlation can be recovered through dressing. 

Compared with FMO-VQE and LAS-nuVQE, the defining feature of ClusterVQE is that the partition is based on entanglement structure rather than on chemical fragments or localized active orbitals. This can be advantageous when correlation is localized but not easily described in conventional fragment language. At the same time, the method becomes less chemically transparent, and its success depends strongly on whether mutual-information-based clustering can capture the physically important couplings. For systems with strong long-range entanglement, the approximation may become less reliable.
\subsection{Where these workflows matter chemically}
The workflows in this section are united by a common idea: the best way to reduce quantum cost is often not to alter the ansatz alone, but to reformulate the chemical problem so that the quantum computer is used only where classical methods are least reliable. Orbital-optimized approaches such as ADAPT-VQE-SCF are most attractive when the main difficulty is a poor one-determinant starting point or a rapidly changing orbital picture, as in bond dissociation, open-shell systems, and transition-metal chemistry. Fragmentation and localized-active-space approaches such as FMO-VQE and LAS-nuVQE are most attractive when correlation is spatially localized, allowing the problem to be decomposed into smaller chemically meaningful pieces. ClusterVQE is appealing when entanglement itself appears localized even if an obvious chemical fragmentation is not available.

To summarize the methods discussed in Sections \ref{ansatz_compression} and \ref{chemicall_motivated}, Table \ref{tab:vqe_compact} compares ansatz construction strategy, expected quantum-resource savings, and principal limitations between VQE variants. A central theme is that reductions in circuit depth or parameter count do not necessarily lead to lower overall cost, since many approaches shift the computational workload to repeated measurements, classical preprocessing, or fragmentation/orbital approximations. The most appropriate method therefore depends on whether the dominant bottleneck is qubit count, two-qubit gate fidelity, measurement costs, or the need to preserve chemically meaningful structure. As a whole, these methods suggest that the most chemically promising uses of VQE may arise not from a standalone fixed ansatz, but from embedding VQE within workflows already familiar to electronic structure theory. Fragment-based approaches such as FMO-VQE and LAS-nuVQE improve scalability by decomposing the problem into smaller correlated units, whereas orbital optimized strategies such as ADAPT-VQE-SCF improve the description of multireference character. The most appropriate choice depends on whether the dominant challenge is system size, strong local correlation, or the need for balanced orbital optimization. 

As a whole, these developments reinforce a central conclusion of this review: VQE is most likely to become chemically useful as a quantum component embedded within established strong-correlation workflows, rather than as a standalone black-box eigensolver.

  \section{Excited-state methods}
  \label{excited_state_methods}
Many chemically relevant problems, including those in spectroscopy, photochemistry, and excited-state reactivity, require access to electronically excited states. Extending VQE beyond the ground state is therefore essential for chemical relevance, but it introduces several additional challenges: preventing variational collapse to lower-energy states, enforcing or preserving orthogonality, targeting specific states within one manifold, and controlling the additional measurement cost needed to obtain multiple states or response properties. This section organizes several extensions of VQE designed for excited-state calculations into two broad categories: state-specific variational targeting methods and subspace/response-based methods.
    
    \subsection{State-specific variational targeting methods}
State-specific variational targeting methods aim to compute excited states one at a time by modifying the standard VQE objective to prevent collapse to lower-energy states and to enforce orthogonality constraints. A prototypical example is the variational quantum deflation (VQD) algorithm, introduced by Higgott et al.
VQD is an extension of the VQE designed to compute excited-state energies of a Hamiltonian on near-term quantum devices.\cite{Higgott2019variationalquantum} It systematically finds the \textit{k}-th excited state by adding overlap penalty terms to the VQE cost function, enforcing orthogonality between the target state and all previously obtained eigenstates.

In standard VQE, the ground-state energy is obtained by using Eq. \ref{vqe_equation_standard}.
In VQD, the cost function for the 
\textit{k}-th excited state is modified as:
\begin{equation}
    F(\theta_k) = \langle\psi(\theta_k)| \hat{H} |\psi(\theta_k)\rangle + \sum_{i=0}^{k-1} \beta_i | \langle \psi(\theta_k) | \psi(\theta_i)\rangle|^2
    \label{VQD_Eq}
\end{equation}
where $\beta_i$ are manually chosen hyperparameters enforcing orthogonality to all previously obtained lower-energy states.

The appeal of VQD is that it extends the familiar VQE optimization framework in a conceptually direct way: each state is still obtained variationally, but overlap penalties discourage collapse back to lower states. VQD maintains the same qubit requirements as VQE and typically doubles the circuit depth due to overlap estimation.
A primary challenge of VQD is sequential error mitigation. Errors in lower states can propagate upward, and the quality of the result depends on choosing penalty weights large enough to enforce orthogonality without making optimization unstable. VQD is therefore most attractive when only a few low-lying states are needed and when one is willing to optimize them one at a time.

Gocho et al. developed an excited-state quantum algorithm that combines VQE with automatically-adjusted constraints (VQE/AC) and a spin-restricted ansatz to improve accuracy and numerical stability on NISQ devices.\cite{gocho2023excited}
The VQE/AC method addresses one of the main practical weaknesses of the VQD approach by eliminating the need to predefine constraint weights ($\beta$ parameters) and instead employing the COBYLA optimization algorithm to automatically enforce orthogonality to lower states. 
This adaptive constraint handling enables smooth potential energy surfaces (PESs) without manual tuning of hyperparameters for every molecular geometry.

A further advantage of the reported implementation is its use of a spin-restricted ansatz. By restricting variational search within the subspace of a given spin multiplicity, the method avoids part of the spin contamination that can otherwise arise in noisy excited-state calculations and can also reduce circuit depth and parameter count relative to more generic ansatz.

The approach was benchmarked on photofunctional molecules like ethylene and phenol blue, specifically focusing on critical geometries such as the Franck-Condon (FC) and conical intersection (CI) points.  The method reproduced ground and excited-state energies with errors below 0.5 kcal mol$^{-1}$ on noisy simulators and within 2 kcal mol$^{-1}$ on IBM's \textit{ibm\_kawasaki} device.
Conceptually, VQE/AC belongs in the same family as VQD: it remains a state-specific constrained optimization scheme. Its main benefit is smoother, more automated orthogonality handling, which is particularly useful along potential-energy surfaces where manual retuning of penalty parameters would otherwise be cumbersome. The tradeoff is that the constrained classical optimization becomes more involved, especially as more excited states are included.
     
Folded spectrum (FS)-VQE has also been proposed as an extension of the standard VQE framework to compute the molecular excited states.\cite{cadi2024folded} Compared to conventional VQE, FS-VQE takes a different route to state-specific targeting. Instead of enforcing orthogonality to lower states, it reformulates the objective so that the desired state is selected by proximity to a target energy $\omega$. The cost function is defined as 
    \begin{equation}
        F = \langle\psi(\theta)\ | (\hat{H} - \omega)^2 | \psi(\theta)\rangle 
    \end{equation}
    and its minimization is carried out variationally within the VQE framework. 
By varying the parameter $\omega$ across the energy spectrum, different excited states can be systematically obtained.

The main price of this direct targeting is measurement cost. Squaring the shifted Hamiltonian dramatically increases the number of Pauli terms that must be measured, so FS-VQE is typically much more expensive in sampling effort than standard VQE-based excited-state methods. 
Furthermore, selecting the target energy $\omega$ presents a practical challenge: if the spectrum is unknown, targeting a specific state requires modifying $\omega$ over a wide range of energies. 

In potential-energy-surface applications this can be eased by state tracking, but the method still relies on having a good estimate of where in the spectrum the state of interest lies. FS-VQE is therefore most appealing when one needs to target a specific interior state and wishes to avoid the sequential error accumulation of deflation-based approaches, provided the increased measurement overhead is acceptable.

\subsection{Subspace and response-based methods}

The subspace-search variational quantum eigensolver (SS-VQE) provides an alternative route to excited-state calculations that avoids the explicit use of overlap penalties or constraint optimization.\cite{nakanishi2019subspace} Instead of enforcing orthogonality between states during the optimization, SS-VQE exploits the fact that orthogonality is preserved under unitary transformations.
In this approach, a set of mutually orthogonal input states \{$\vert\varphi
_j$$\rangle$\}, typically chosen from computational basis states, is propagated through a parameterized unitary ansatz $U(\theta)$. 
The optimization is performed by minimizing a cost function defined over the entire subspace:
\begin{equation}
    L(\theta)  = \sum_{j=0}^k \langle \varphi_j \vert U^\dagger(\theta)HU(\theta)\vert\varphi_j\rangle
\end{equation}
This procedure variationally identifies a low-energy subspace that approximates the span of the lowest $k$+1 eigenstates of the Hamiltonian. Because the input states are orthogonal and the ansatz is unitary, the resulting output states remain orthogonal by construction, eliminating the need for overlap measurements or swap-test circuits.

In this sense, SS-VQE is not a state-by-state method but a low-energy-subspace method: it first identifies a variationally optimized manifold and then extracts states within that manifold. From a practical standpoint, SS-VQE is particularly attractive for NISQ devices because it avoids ancilla qubits and overlap estimation circuits, leading to shallower implementations compared to penalty-based methods such as VQD. 
Its main challenge is that the ansatz must be expressive enough to represent several states simultaneously, which can complicate optimization relative to state-specific targeting. SS-VQE is therefore most attractive when several low-lying states are needed together, as in spectroscopy or low-energy valence manifolds, rather than when one specific state is the sole target.

\begin{table*}[!htb] 
\caption{Qualitative comparison of the excited state VQE variants discussed in Sections \ref{excited_state_methods}. } \label{tab:excited_state}
\centering \scriptsize \setlength{\tabcolsep}{4pt} \renewcommand{\arraystretch}{1.1} \begin{tabular}{|p{1.8cm}| p{2.8cm}| p{4.0cm} |p{4.0cm}| p{2.9cm}|} 
\hline 
\textbf{Method} & \textbf{Type} & \textbf{Main benefit} & \textbf{Main cost} & \textbf{Typical niche} \\ \hline 
VQD & Deflation-based state-specific VQE &  Simple VQE extension; explicit orthogonality & Overlap measurements, penalty-weight tuning, and sequential error accumulation & A few low-lying excited states computed one at a time\\\hline 
VQE/AC & Constrained state-specific VQE & No manual penalty tuning; smoother PESs; can use spin-restricted ansatz & More involved constrained classical optimization & Excited-state PESs and  photochemistry \\\hline 
FS-VQE & Folded-spectrum state targeting & Direct targeting of interior states  & Large measurement overhead & Specific interior excited states \\\hline 
SS-VQE & Variational subspace method & Orthogonality by construction; multiple states at once & Harder subspace optimization; expressive ansatz needed & Low-lying manifolds and multistate spectroscopy\\\hline 
qEOM & Response/equation-of-motion method & Multiple excitation energies from a single VQE ground-state reference & Large measurement overhead from response matrices and higher-order RDMs & Small multistate excited-state calculations \\\hline 
mcEOM &  Multicomponent EOM method
&Treats coupled excitations of multiple particle types
& Large secular space; high measurement cost
& Non-Born-Oppenheimer and multicomponent chemistry\\\hline 
q-sc-EOM & Self-consistent EOM method
& Lower measurement cost; access to EE, IP, and EA
& Depends on reference quality and operator manifold
&Resource-aware EE/IP/EA calculations\\\hline 
oo-VQE-qEOM&
 Orbital-optimized EOM method
& Includes orbital relaxation and transition properties
& More elaborate hybrid workflow 
& Spectroscopy and active-space chemistry\\
\hline
\end{tabular}
\end{table*}
A second major family of excited-state methods is inspired by classical linear-response and equation-of-Motion (EOM) theory. Rather than optimizing each excited state variationally, these methods begin from an optimized ground-state wavefunction and obtain multiple excitation energies by constructing and solving an effective eigenvalue problem in a space of excitations built on top of that reference. 

The quantum Equation-of-Motion (qEOM), introduced by Ollitrault et al., adapts the classical EOM formalism to the constraints of a hybrid quantum-classical computing environment, offering an efficient extension of the VQE method for calculating molecular excitation energies.\cite{ollitrault2020quantum}
The theoretical foundation for calculating the excitation energy,  
        $E_{0n} = E_n-E_0$
 , is derived from the double commutator, which ensures the resulting energy differences are real and the operators are Hermitian. The core expectation value for the excitation energy is thus given by:
 \begin{equation}
     E_{0n} = \frac{\langle0|[\hat{O}_n, [\hat{H},\hat{O_n^\dagger}]]|0\rangle}{\langle0|[\hat{O}_n,\hat{O}_n^\dagger]|0\rangle}
 \end{equation}
 The excitation operator $\hat{O}_n^\dagger$ is systematically expanded as a linear combination of basis excitation ($\hat{E}$) and deexcitation ($\hat{E}^\dagger$) operators, typically restricted to single and double electronic excitations for practical quantum implementation
 \begin{equation}
     \hat{O_n^\dagger} = \sum_\mu [X^{(\alpha)}_{\mu_\alpha}(n)\hat{E}^{(\alpha)}_{\mu_\alpha} -Y^{(\alpha)}_{\mu_\alpha}(n)( \hat{E}^{(\alpha)}_{\mu_\alpha})^\dagger] 
 \end{equation}
 %Importantly, the inclusion of the deexcitation coefficients (Y) is what differentiates qEOM from methods based on the Tamm-Dancoff approximation, making it applicable to systems exhibiting strong correlation. 
 Applying the variational principle to this formulation leads directly to the core mathematical challenge: a generalized secular equation that must be solved classically:
 \begin{equation}
     \begin{pmatrix}
         M & Q \\
         Q^* & M^* \\
     \end{pmatrix}
     \begin{pmatrix}
         X_n
         \\
         Y_n\\ 
     \end{pmatrix}
     = E_{0n} 
     \begin{pmatrix}
         V & W \\
         -W^* & -V^* \\
     \end{pmatrix}
     \begin{pmatrix}
         X_n
         \\
         Y_n\\ 
     \end{pmatrix}
 \end{equation}

The practical structure is straightforward: first optimize a ground-state VQE reference, then measure the matrix elements needed for the EOM secular problem, and finally solve the resulting eigenvalue problem classically. Compared with state-specific methods, the advantage is that one ground-state optimization can yield a manifold of excited states. The main limitation is measurement cost, since the required matrix elements can involve high-order reduced density matrices. 

For systems where the Born-Oppenheimer approximation is invalid or insufficient, such as those involving photoinduced proton transfer or light particles like positrons, the multicomponent Equation-of-Motion (mcEOM) method extends this framework.\cite{pavosevic2021multicomponent} The mcEOM, paired with the Multicomponent Unitary Coupled Cluster (mcUCC) ansatz, allows for the quantum mechanical treatment of multiple particle types (e.g., electrons and nuclei/positrons) simultaneously. This is achieved by formulating the excitation operator  
$\hat{O_n^\dagger}$
to include both electronic and non-electronic particle excitations, leading to a secular equation identical in form to the standard EOM-CC method, but with much greater dimensionality due to the inclusion of all particle types. The resulting algorithm, mcEOM-VQE, is crucial for studying non-Born-Oppenheimer processes, where excitations can involve the simultaneous transition of electrons and quantum nuclei.
  
A further development is the quantum self-consistent Equation-of-Motion (q-sc-EOM) method.\cite{asthana2023quantum} q-sc-EOM was introduced to address both conceptual and practical weaknesses of qEOM by enforcing the vacuum annihilation condition self-consistently. 
  
It defines the state-transfer operator $\hat{O_k}$ such that 
  \begin{equation}
      \hat{O_k}|\psi_{gr}\rangle = |\psi_k\rangle,
  \end{equation}
  and by using self-consistent operators to enforce VAC i.e., $\hat{O_k^\dagger}|\psi_{gr}\rangle = 0$.
  By rigorously satisfying the VAC, q-sc-EOM provides highly accurate and guaranteed real energy differences, making it suitable for calculating electronic excitation energies, IPs, and EAs.
  %, which qEOM fails to do accurately.
  The overlap matrix $V$ becomes the identity matrix and the off-diagonal matrices $Q$ and $W$ vanish, simplifying the problem to a standard, Hermitian eigenvalue problem, 
  \begin{equation}
      MA_k = E_{0k}VA_k
  \end{equation}
  
As a result, q-sc-EOM reduces measurement demands substantially relative to qEOM and extends naturally to properties such as ionization potentials and electron affinities. It is therefore one of the most practically attractive EOM variants when one wants a broader set of state-to-state or particle-changing observables without the full cost of higher-order RDM measurements.
  
To address the challenge of simulating systems with large basis sets on resource-limited hardware, the orbital-optimized-VQE - quantum Equation-of-Motion (oo-VQE-qEOM) protocol was introduced.\cite{jensen2024quantum} This method combines active-space/orbital optimization with an EOM treatment of excitations, making it the closest quantum analogue to classical multiconfigurational response methods such as CASSCF-EOM. The core strategy is the Active Space Approximation, which partitions the molecular orbitals into three sets: inactive (always occupied), virtual (always unoccupied), and active (correlated on the QPU). The active-space approximation has the advantage that the qubits corresponding to inactive and virtual parts can be removed, and only the active-space will be simulated on a quantum computer reducing the number of qubits and gates needed for the simulation. The Hamiltonian is then restricted to the active-space, as shown:
  \begin{equation}
      \hat{H}(\vec{\kappa}) = \sum_{pq}^N h_{pq}(\vec{\kappa})\hat{E}_{pq} + \frac{1}{2} \sum_{pqrs}^N g_{pqrs}(\vec{\kappa})(\hat{E}_pq\hat{E}_rs - \delta_{qr}\hat{E}_ps) + h_{nuc}
  \end{equation}

The main appeal of oo-VQE-qEOM is that it combines three features especially relevant for chemistry: active-space compression, orbital relaxation, and access not only to excitation energies but also to response properties such as oscillator and rotational strengths. It is therefore particularly promising for spectroscopy-oriented applications where chemically meaningful observables extend beyond state energies alone.

To summarize, the subspace/response-based methods offer a different tradeoff from state-specific approaches. They avoid repeated full variational optimizations for each state and naturally provide a manifold of excitations referenced to a common ground state, which is attractive for spectroscopy and multistate chemistry. Their main cost is that one must measure the matrix elements or reduced density matrices needed to define the effective excited-state problem. Within this family, SS-VQE keeps the problem variational and explicitly subspace-based, qEOM and mcEOM emphasize response theory built on a VQE reference, q-sc-EOM reduces the measurement burden through self-consistency, and oo-VQE-qEOM adds orbital relaxation and access to spectroscopic observables. Table \ref{tab:excited_state} summarizes these distinctions.

\subsection{Chemistry applications of the excited-state methods}
The methods reviewed above are best understood not as direct competitors in all settings, but as tools suited to different excited-state tasks. Table \ref{tab:excited_state} briefly summarizes the discussed excited-state methods. State-specific methods such as VQD and VQE/AC are most natural when only a few low-lying states are needed and when one wants to optimize each state directly. VQE/AC is especially attractive when orthogonality constraints must be enforced smoothly along a potential-energy surface, while VQD remains appealing as the simplest conceptual extension of ground-state VQE. FS-VQE is best viewed as a specialized targeting method for interior states, useful when one wants to avoid the sequential dependence of deflation-based schemes and is willing to pay a substantially higher measurement cost.

Subspace and response-based methods are more attractive when the interest is a manifold of states rather than a single excitation. SS-VQE is well suited to cases where one wants several low-energy states within one optimized variational framework, while EOM-based methods are particularly appealing when many excitation energies are needed from a single correlated ground state. Among the EOM family, q-sc-EOM is especially promising when reduced measurement cost and access to IP/EA sectors matter, whereas oo-VQE-qEOM is the most chemically complete option when orbital relaxation and spectroscopic observables such as oscillator strengths are important. mcEOM is more specialized, but it points toward applications in non-Born–Oppenheimer excited-state chemistry where multiple particle types must be treated quantum mechanically.

For chemistry, this suggests a practical division of labor. Few-state photochemical problems at fixed active-space may favor VQD or VQE/AC. Multistate spectroscopy and low-lying manifolds are more naturally matched to SS-VQE or EOM-like methods. Applications requiring transition properties or larger basis descriptions through active-space compression are better aligned with oo-VQE-qEOM. Across all cases, the central tradeoff remains the same: methods that simplify orthogonality or state targeting often increase measurement burden, while methods that obtain many states from one reference shift the cost toward response-matrix construction and reduced-density-matrix measurement.

\section{Benchmarking}
\label{benchmarking}
The calculations in this section are original, illustrative benchmarks performed for this paper and are intended to illustrate, rather than definitively rank, how different ansatz construction strategies redistribute cost across accuracy, parameter count, excitation structure, and expected measurement overhead. To make this comparison concrete, we focus on ADAPT-VQE and USCC as two contrasting approaches to ansatz compression: ADAPT-VQE builds the wavefunction adaptively through repeated gradient evaluation on quantum hardware, whereas USCC relies on classical prescreening and selective inclusion of higher-order disconnected excitations. Both are compared against standard UCCSD-VQE and representative classical coupled-cluster methods, using FCI within the same basis as the reference. 

The primary metrics are energy error relative to FCI, final ansatz size and excitation-operator composition, and operator-pool size used as a proxy for adaptive measurement overhead. Explicit two-qubit gate counts are not uniformly available for all implementations considered here and are therefore discussed qualitatively rather than quantitatively. The present benchmarks are restricted to noiseless simulation and therefore do not capture the tradeoff between shallower circuits and repeated measurements under realistic device noise. The computational details for each method used in the benchmarking study are given in the Supporting Information. 

Method development and benchmarking in this area are often carried out in noiseless statevector or tensor-network simulators, which provide a controlled setting for isolating ansatz design tradeoffs, convergence behavior, and measurement-scaling trends without conflating these effects with hardware noise. This practice is especially useful at the present stage of the field, where comparisons are often intended to clarify algorithmic tradeoffs rather than to reproduce device-specific performance. For convenience to readers less familiar with the simulation software available and commonly used in quantum chemistry benchmarking, we summarize in the Supporting Information the quantum simulators and qubit-mapping software frameworks most frequently used in the literature discussed in this paper.

\subsection{Accuracy across potential-energy surfaces}
\begin{figure*}[h!]
\caption{RHF, CCSD, CCSD(T), UCCSD(T), VQE and ADAPT-VQE (threshold $\epsilon^{-4}$) energy differences from FCI plotted on a log scale for increasing interatomic distances with the STO-3G minimal basis set. The molecules tested are A. non-symmetric H$_2$O B. non-symmetric BeH$_2$ and C. non-symmetric H$_6$. The  shaded area represents data points below chemical accuracy. }
\centering
\includegraphics[width=1.0\textwidth]{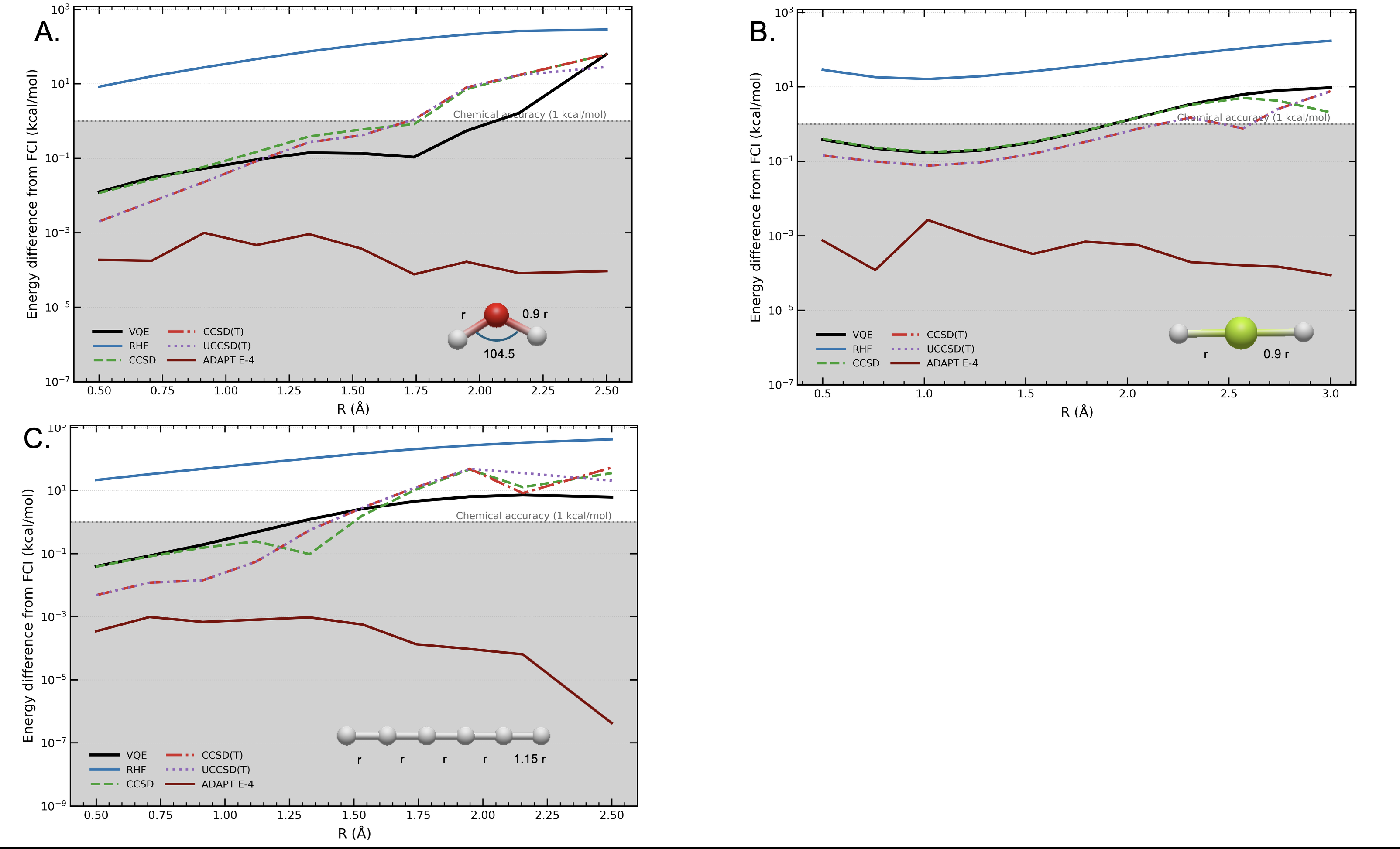}
\label{fig:1}
\end{figure*}

Energies for restricted Hartree-Fock (RHF), coupled cluster with singles and doubles (CCSD) and non-iterative triples (CCSD(T)), and unrestricted CCSD(T) (UCCSD(T)) were calculated with the STO-3G basis set and errors against full configuration interaction (FCI) are reported in log scale for non-symmetric H$_2$O, non-symmetric BeH$_2$, and non-symmetric H$_6$ in Figure \ref{fig:1}. Errors for both standard UCCSD-VQE and ADAPT-VQE are plotted to compare the accuracy of these methods with that of classical methods including the so-called "gold standard" classical method CCSD(T). Both restricted and unrestricted CCSD(T) are included to capture important information about potential unpaired electrons in the large bond stretching regions where energy errors often occur for methods built upon restricted HF methods. Only the largest values of R show any divergence between CCSD(T) and UCCSD(T). As expected, the baseline method RHF has high errors well above chemical accuracy. In the bonding regions around equilibrium all other methods have errors below chemical accuracy. Across the full potential energy surface (PES) for each molecule, there is little deviation between CCSD, CCSD(T), UCCSD(T), and standard VQE due to the underlying excitation schemes being the same between each one. ADAPT-VQE with a tight cutoff shows smaller errors than the other methods considered across the full PES for these molecules.

\begin{figure*}[h!]
\caption{ADAPT-VQE, USCC, and UCCSD-VQE energy differences from FCI plotted on a log scale for increasing interatomic distances with thresholds values of $\epsilon^{-1}$, $\epsilon^{-2}$, $\epsilon^{-3}$, and $\epsilon^{-4}$. The molecules tested are A. H$_2$, B. LiH, C. non-symmetric H$_6$, D. non-symmetric H$_2$O, and E. non-symmetric BeH$_2$. The minimal STO-3G basis set was used for each calculation. The  shaded area represents data points below chemical accuracy. }
\centering
\includegraphics[width=1.0\textwidth]{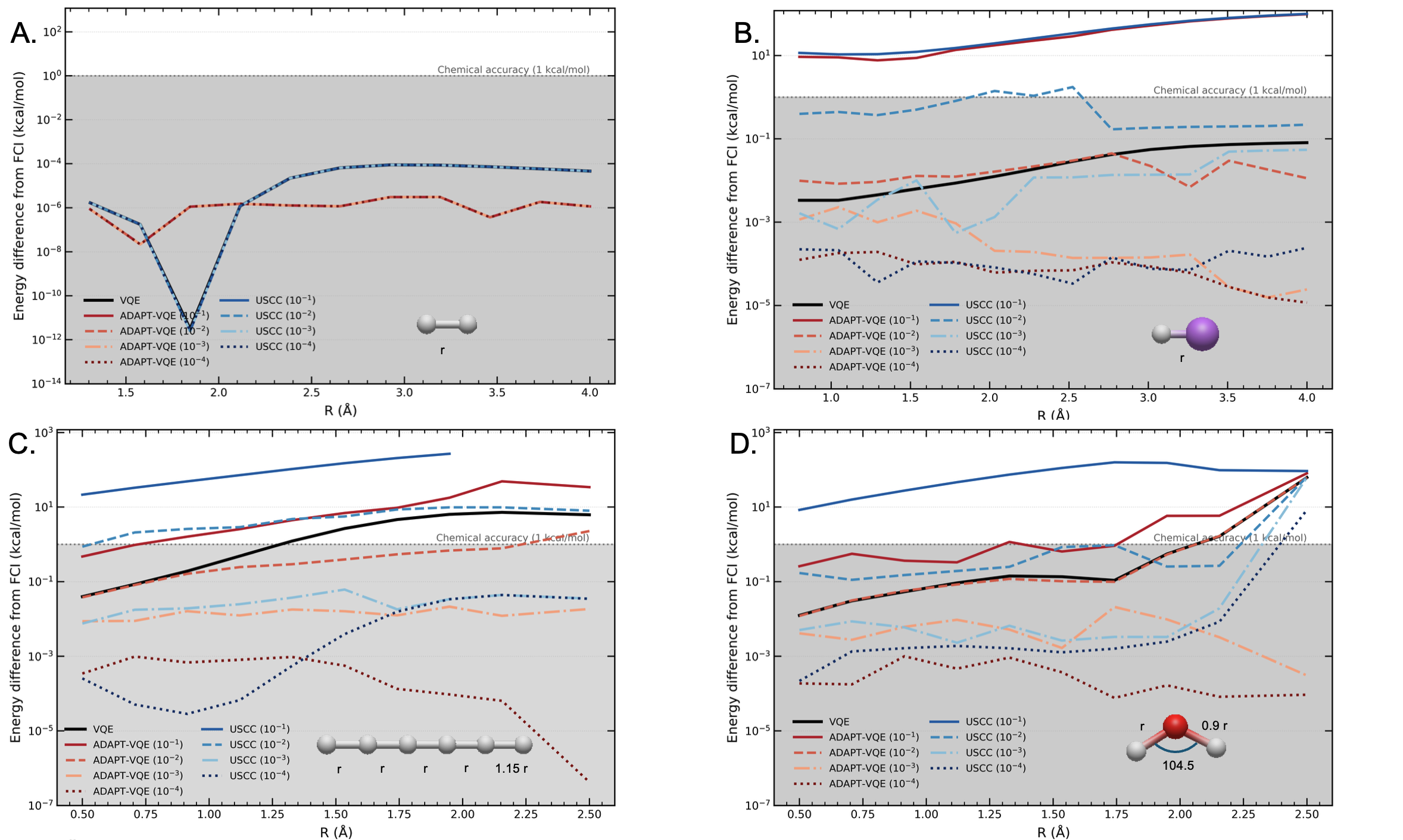}
\label{fig:2}
\end{figure*}

ADAPT-VQE and USCC are then compared to standard VQE with varying threshold values for accuracy compared to FCI and quantum resource requirements. Figure \ref{fig:2} provides the accuracy of varying threshold values against FCI. Both ADAPT-VQE and USCC were run with thresholds of $\epsilon^{-1}$, $\epsilon^{-2}$, $\epsilon^{-3}$, and $\epsilon^{-4}$. For ADAPT-VQE with the STO-3G basis set, a threshold value of $\epsilon^{-2}$ will achieve chemical accuracy across the PES for all but the largest bond stretches and a threshold value of $\epsilon^{-3}$ will produce errors below those of standard VQE across the PES. In the case of USCC, the threshold $\epsilon^{-3}$ is required to achieve errors below chemical accuracy across the full PES for all but the largest bond stretches and for most of the molecules this threshold is enough to have smaller errors than standard VQE. The PES for BeH$_2$ requires a threshold cutoff of $\epsilon^{-4}$ for USCC to be more accurate than standard VQE across the PES. The exception to the threshold values is the H$_2$ molecule where USCC for all thresholds are equivalent to standard VQE and all ADAPT-VQE thresholds generate the same errors. The small number of potential excitations for H$_2$ with the minimal basis set does not allow for a sufficient number of operators in the generated operator pool to distinguish between the various methods and threshold values. 

Based on the benchmark systems considered here, we found that for ADAPT-VQE a threshold value of $\epsilon^{-2}$ will generate errors below chemical accuracy and $\epsilon^{-3}$ will produce more accurate results than standard VQE. Similarly, for USCC threshold values of $\epsilon^{-3}$ and $\epsilon^{-4}$ are required to produce accuracy below chemical accuracy and standard VQE respectively. At any point on the PES of the test molecules both ADAPT-VQE and USCC are capable of accuracy below that of standard VQE with a tight enough threshold. For the tight threshold of $\epsilon^{-4}$ both USCC and ADAPT-VQE have similar accuracies, except for at large bond stretching distances. 

\subsection{Parameter count and excitation-rank composition}
The improved accuracy of both USCC and ADAPT-VQE relative to standard UCCSD-VQE reflects the fact that both depart from a fixed occupied-to-virtual excitation pattern, but via different ways. As described in Section \ref{ansatz_compression}, USCC utilizes disconnected triple and quadruple excitations in addition to the connected single and double excitations utilized in standard VQE. The larger excitation schemes allow USCC to capture more correlation energy that can only be captured by higher-order excitations. This comes at the cost of an increase in the number of parameters, sometimes well above that of standard VQE. ADAPT-VQE, on the other hand, generates a larger number of excitations by utilizing the generalized CCSD ansatz as described in Section 2.1. This excitation scheme does not distinguish between occupied and virtual orbitals when constructing excitation operators allowing for the potential of disconnected higher order excitations to be included. This difference becomes clear when the final ansatz sizes and excitation-rank compositions are examined.

\begin{figure*}[h!]
\raggedright
\includegraphics[width=0.47\textwidth]{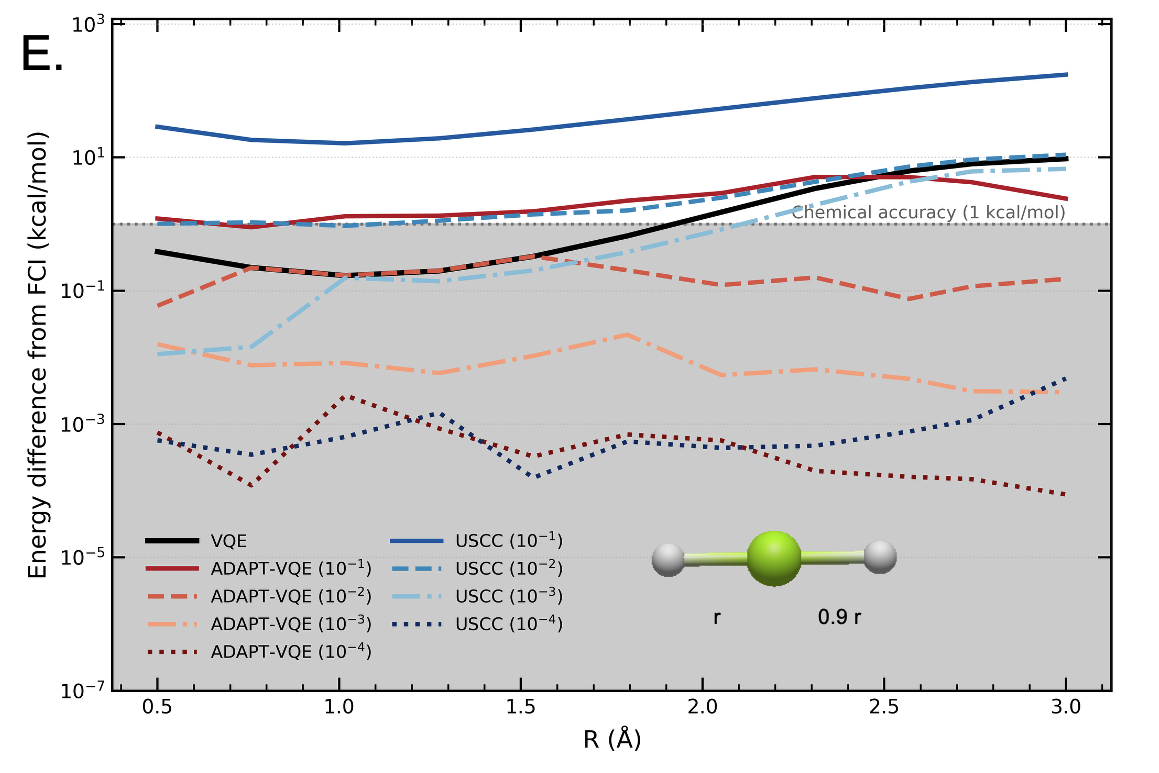}
\end{figure*}

When comparing circuit costs, neither ADAPT-VQE nor USCC calculations natively provide two-qubit gate counts without code modification and therefore we do not report explicit two-qubit gate counts for these calculations. Instead we report the number and rank of operators in each ansatz to provide a meaningful proxy for circuit resource requirements. Under standard fermion-to-qubit mappings each excitation operator decomposes into a circuit block whose CNOT count depends on the excitation rank (single, double, triple, etc.). Higher-rank excitations require more two-qubit gates per operator than lower-rank excitations. Consequently, while the total parameter count does not map to a unique gate count, ansatzes with more parameters, and especially those like USCC that contain higher-order disconnected excitations, will generally require deeper circuits with more two-qubit gates. Breaking down the operator composition by excitation rank therefore provides additional insight into the expected circuit cost beyond the total parameter count alone. 

Table \ref{table:1} shows the average number of parameters required for non-symmetric H$_2$O (STO-3G) basis and linear H$-4$ (STO-3G and 6-31G) at each threshold value for USCC and ADAPT-VQE compared to standard VQE. The number of parameters for the other molecules is given in the Supporting Information (Table S1-S5).  While ADAPT-VQE and USCC can generate similar errors compared to FCI for similar thresholds, the number of parameters required to achieve that accuracy are quite different. In order to match ADAPT-VQE's accuracy with threshold values of $\epsilon^{-4}$, USCC requires over 2.6X the number of parameters. Similar trends occur for other molecules: 2.7X for LiH and non-symmetric H$_6$, and 3.4X for BeH$_2$. The average parameter counts can be further broken down into the average number of single, double, etc. excitation operators as given in Table \ref{table:2}. The average number of single, double, etc. for the other molecules is given in the Supporting Information (Table S6). As the threshold value tightens the number of disconnected higher-order excitation operators for USCC quickly increases. Compared to the generalized excitation operators for ADAPT-VQE, USCC relies on larger number disconnected triple and quadruple excitation operators to increase the accuracy of the method. These disconnected triple and quadruple excitation operators require more two-qubit gates to be implemented on quantum circuits. By using the generalized excitation scheme to generate the pool of operators, ADAPT-VQE can achieve similar or lower errors than USCC with fewer parameters. Further, ADAPT-VQE operators correspond only to single or double excitations, which do not require the deeper circuits associated with the higher-order excitations used in USCC.

As mentioned in Section 2.1, ADAPT-VQE must perform gradient evaluations for each operator in the operator pool at each iteration increasing the cost on quantum hardware. Table \ref{table:3} lists the size of the operator pool for each benchmarking system for ADAPT-VQE. USCC does not incur an increasing gradient cost as all the operator pool selections occur through classical prescreening. So while ADAPT-VQE can achieve similar or better accuracies than USCC compared to FCI with shallower circuit requirements, this does come at the cost of increasing numbers of gradient evaluations as active-space size grows.

\begin{table*}[t] \centering \caption{Average number of parameters for ADAPT-VQE, USCC, and standard VQE at different threshold values for non-symmetric H$_2$O with the STO-3G basis set and symmetric H$_4$ with the STO-3G and 6-31G basis sets.} \label{table:1} \footnotesize \renewcommand{\arraystretch}{1.15} \begin{tabular*}{\textwidth}{@{\extracolsep{\fill}} l c c c} \hline \textbf{Method} & \textbf{H$_4$ (STO-3G)} & \textbf{H$_4$ (6-31G)} & \textbf{H$_2$O (STO-3G)} \\ \hline VQE & 26 & 198 & 140 \\ ADAPT-VQE ($\epsilon^{-1}$) & 5 & 26 & 10 \\ ADAPT-VQE ($\epsilon^{-2}$) & 9 & 48 & 25 \\ ADAPT-VQE ($\epsilon^{-3}$) & 10 & 76 & 50 \\ ADAPT-VQE ($\epsilon^{-4}$) & 11 & 107 & 75 \\ USCC ($\epsilon^{-1}$) & 4 & 7 & 8 \\ USCC ($\epsilon^{-2}$) & 14 & 74 & 36 \\ USCC ($\epsilon^{-3}$) & 19 & 254 & 123 \\ USCC ($\epsilon^{-4}$) & 19 & 392 & 199 \\ \hline \end{tabular*} \end{table*}

\begin{table*}[t] \centering \caption{Average number of single (s), double (d), triple (t), and quadruple (q) excitation operators for ADAPT-VQE (s,d), USCC (s,d,t,q), and standard VQE (s,d) at different threshold values for non-symmetric H$_2$O with the STO-3G basis set and symmetric H$_4$ with the STO-3G and 6-31G basis sets.} \label{table:2} \footnotesize \renewcommand{\arraystretch}{1.15} \begin{tabular*}{\textwidth}{@{\extracolsep{\fill}} l c c c} \hline \textbf{Method} & \textbf{H$_4$ (STO-3G)} & \textbf{H$_4$ (6-31G)} & \textbf{H$_2$O (STO-3G)} \\ \hline VQE & 8, 18 & 24, 174 & 20, 120 \\ ADAPT-VQE ($\epsilon^{-1}$) & 0, 5 & 1, 25 & 0, 10 \\ ADAPT-VQE ($\epsilon^{-2}$) & 0, 9 & 1, 47 & 0, 25 \\ ADAPT-VQE ($\epsilon^{-3}$) & 0, 10 & 2, 72 & 0, 50 \\ ADAPT-VQE ($\epsilon^{-4}$) & 0, 11 & 2, 105 & 0, 75 \\ USCC ($\epsilon^{-1}$) & 4, 0, 0, 0 & 7, 0, 0, 0 & 8, 0, 0, 0 \\ USCC ($\epsilon^{-2}$) & 4, 10, 0, 0 & 11, 63, 0, 0 & 12, 22, 1, 1 \\ USCC ($\epsilon^{-3}$) & 4, 10, 4, 1 & 12, 89, 124, 29 & 15, 47, 51, 10 \\ USCC ($\epsilon^{-4}$) & 4, 10, 4, 1 & 12, 90, 179, 111 & 15, 66, 88, 30 \\ \hline \end{tabular*} \end{table*}

\begin{table*}[!ht] \caption{Operator pool size for each molecule at STO-3G (unless otherwise noted) for ADAPT-VQE to demonstrate the gradient evaluation requirements.} \label{table:3} \centering \footnotesize \renewcommand{\arraystretch}{1.15} \begin{tabular*}{\textwidth}{@{\extracolsep{\fill}} l c c c c c c c} \hline  & \textbf{H$_2$} & \textbf{LiH} & \textbf{BeH$_2$} & \textbf{H$_2$O} & \textbf{H$_4$} & \textbf{H$_4$ (6-31G)} & \textbf{H$_6$} \\ \hline Operator Pool Size & 4 & 330 & 609 & 609 & 66 & 1036 & 330 \\ \hline \end{tabular*} \end{table*}

\subsection{Basis-set dependence of threshold strategies}
\begin{figure*}[h!]
\caption{ADAPT-VQE, USCC, and VQE energy differences from FCI plotted on a log scale for increasing interatomic distances for symmetric H$_4$ on both A. the minimal basis set STO-3G and B. a larger basis set 6-31G to highlight any basis set dependence. As before the thresholds values were set to $\epsilon^{-1}$, $\epsilon^{-2}$, $\epsilon^{-3}$, and $\epsilon^{-4}$. Each collection of results with a given basis set is compared to the FCI energy calculated with that same basis set. The USCC $\epsilon^{-3}$, and $\epsilon^{-4}$ data points for A. are overlapping and only show the one line trend. The  shaded area represents data points below chemical accuracy. }
\centering
\includegraphics[width=1.0\textwidth]{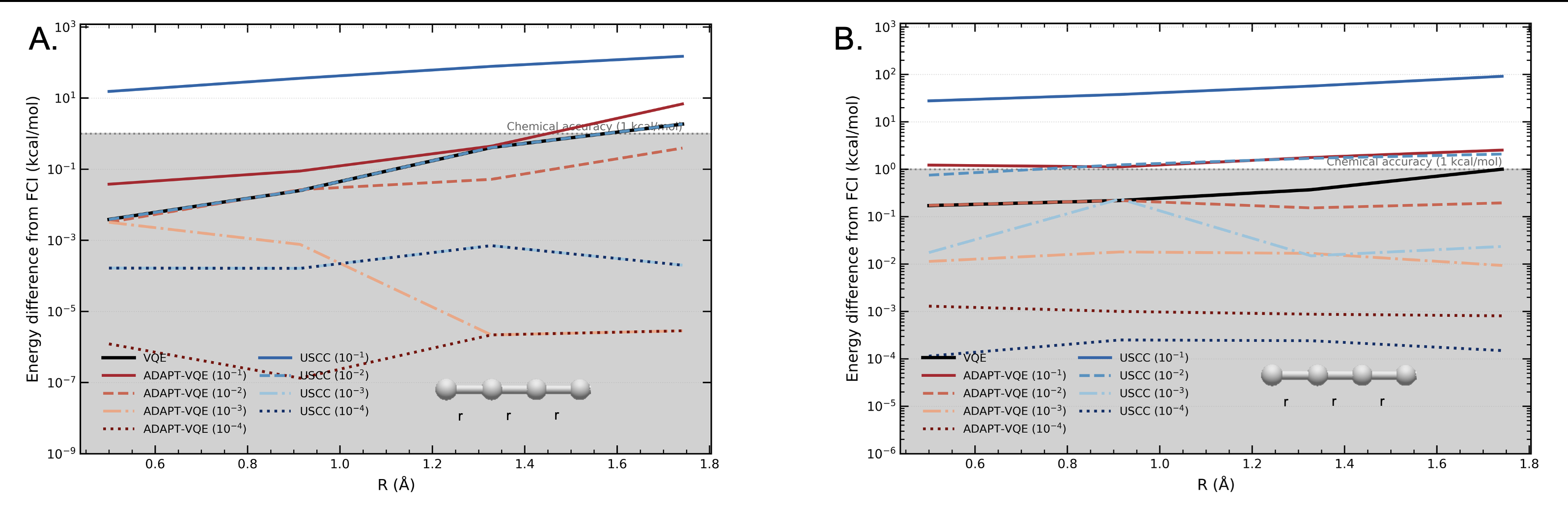}
\label{fig:3}
\end{figure*}

To test the effect of different basis sets on the accuracy of each ansatz design, calculations were performed on a symmetric H$_4$ with both the minimal STO-3G basis set and the 6-31G Pople basis set. The difference in errors for each different method is shown in Figure \ref{fig:3} highlighting the basis set dependence of each method. The reported energy differences are calculated with respect to the FCI energy at the given basis set (VQE with STO-3G is compared with FCI at STO-3G, etc.).

For all threshold values for both USCC and ADAPT-VQE, with the exception of USCC with $\epsilon^{-4}$, the error between the VQE and FCI energy was larger for 6-31G compared to that for STO-3G. The number of parameters required by USCC to match the accuracy of ADAPT-VQE with $\epsilon^{-4}$ also increases with larger basis set, from 1.7X to 3.7X parameters for STO-3G and 6-31G, respectively. Further, there are larger error differentiations among different $\epsilon$ values for the bigger basis set. This is highlighted most clearly by the overlapped error data points between thresholds $\epsilon^{-3}$ and $\epsilon^{-4}$ for each method (Figure \ref{fig:3}A), which are more clearly differentiated in the 6-31G basis set plot (Figure \ref{fig:3}B). 

The changes in absolute errors and the distinctions introduced between threshold values are likely due to the increase in the number of available excitations from a larger number of virtual orbitals in the larger basis set. Larger basis sets do not expand the number of occupied orbitals, but expand the space of virtual orbitals available to the calculation. With a larger number of virtual orbitals, the number of possible excitations in the operator pool increases for both ADAPT-VQE and USCC. With the minimal basis set, the possible number of operators for a small molecule is limited, leading to a smaller set of operators with large gradient contributions. This means that a lower threshold value may not be able to capture any more meaningful correlation energy which leads to the lack of distinction between the threshold values. When utilizing a larger basis set, which generates a larger operator pool, the number of operators that have small contributions to the correlation energy increase. With the increase in the number of operators with small gradient contributions, larger distinctions arise among different threshold values. 

The increase in the number of operators with smaller gradient contributions creates the basis set dependence of these adaptive methods. As the basis set size increases, the FCI energy gets closer to the exact energy (i.e. progressing towards complete basis set extrapolation). As this virtual orbital space increases (and therefore the VQE operator pools increase), the cutoff thresholds capture a smaller percentage of these low-lying contributions to the correlation energy, leading to lower accuracy of the VQE methods with respect to the FCI energy. Therefore, as basis-set size increases the cutoff threshold values must also be adjusted to capture an equivalent amount of information as compared to FCI. 

\section{Conclusions}
\label{conclusions}
The VQE method has emerged as the leading framework for near-term quantum chemistry because it replaces the deep coherent circuits required by fault-tolerant algorithms with a hybrid workflow based on state preparation, measurement, and classical optimization. This shift makes VQE far more compatible with current hardware, but it also means that progress depends not only on circuit design, but on how computational cost is redistributed across measurements, classical preprocessing, and chemically motivated approximations. 

In response to the limitations of standard VQE, recent developments have increasingly emphasized resource-aware and chemically structured workflows. Across the methods reviewed here, a clear theme is that no single ansatz or workflow is uniformly best. Adaptive strategies such as ADAPT-VQE can produce compact and accurate ansatzes, often with better potential-energy-surface behavior than fixed UCCSD-like forms, but they pay for this compactness through repeated gradient measurements. Selective approaches such as USCC reduce the quantum cost of ansatz construction by shifting screening to the classical computer, but may require larger final operator sets and higher-rank excitations. Non-unitary extensions such as nuVQE improve expressivity without increasing circuit depth, but leave the measurement burden largely intact. 

Our illustrative benchmarks reinforce this broader point: savings in parameter count do not automatically imply savings in total computational cost, and threshold-based conclusions are strongly basis set dependent.

A second major conclusion is that the most promising advances for chemistry come not only from ansatz compression, but from embedding VQE within workflows already familiar in electronic structure theory. Orbital-optimized approaches can improve the balance and compactness of active-space wavefunctions in multireference settings. Fragmentation and localized active-space methods can reduce qubit requirements dramatically when correlation is spatially structured. Entanglement-based partitioning offers a complementary path when useful decompositions are more naturally defined in qubit space than by chemical fragments. These strategies do not remove approximation; rather, they make it explicit and chemically interpretable.

Excited-state extensions further broaden the scope of VQE beyond ground-state single-reference problems. State-specific methods such as VQD, VQE/AC, and FS-VQE offer direct routes to selected excited states, while subspace and EOM-based methods provide more natural access to manifolds of states and, in some cases, to spectroscopic observables. Here again, the tradeoff is not whether one method is simply “better”, but whether the application favors direct state targeting, smooth multistate balance, or response-style access to several excitations from a common reference.

To summarize, the developments reviewed here suggest that the chemically useful future of VQE is unlikely to be defined by a single universally preferred ansatz. Instead, it will likely emerge from chemistry-aware hybrid workflows that selectively deploy quantum resources on the most strongly correlated or multistate parts of the problem, while using classical structure through orbitals, active spaces, fragments, and response formalisms to control cost and preserve interpretability.

   \section{Perspective}
   \label{perspective}
The broad range of VQE variants discussed above suggests that the most important question is which kinds of chemical problems most justify the quantum resources, approximations, and workflow complexity that VQE entails. In our view, the strongest case for chemically useful VQE does not lie in equilibrium single-reference molecules, where established classical methods already provide high accuracy at lower cost, but in electronically challenging regimes where classical approaches become increasingly delicate, expensive, or method dependent. Two such regimes stand out. The first is multireference ground-state chemistry, particularly in catalytic systems with open shells, near-degeneracy, and large active spaces.\cite{Rishu_2024,Gagliardi_2019,Truhlar_2022,Gagliardi_2025,Gagliardi_2022,khurana2026multireference,Gagliardi_2026} The second is excited-state chemistry, where multiple low-lying states, changing electronic character, and spectroscopic observables must often be treated on equal footing.\cite{Rishu_1,Patel2022,Potter2015,Rishu_2025} These are also the regimes where the recent methodological developments in adaptive ansatz design, orbital optimization, fragmentation, and subspace-based excited-state treatments appear most aligned with genuine chemical need.
\subsection{Multireference electronic structure as a central target for chemistry VQE}
A recurring lesson in both classical and quantum electronic structure theory is that the most chemically difficult problems are often not those with the largest number of atoms, but those in which a qualitatively correct wavefunction requires multiple determinants of comparable importance. This multireference character arises when several orbitals are near degenerate or when electron pairing and orbital occupation change substantially along a reaction coordinate. Examples include bond dissociation, spin-state ordering, diradicals, transition-metal complexes, and catalytic intermediates that undergo redox changes or ligand-field rearrangements. In such systems, single-reference pictures based on a Hartree-Fock determinant can become qualitatively inadequate, and methods that depend strongly on that reference may lose reliability or become difficult to converge systematically.

In our view, this is the most compelling chemistry target for VQE. The appeal is not that VQE should replace classical single-reference methods such as CCSD(T) on routine molecules, but rather that it may provide a flexible variational route to strongly correlated active spaces whose exact or near-exact treatment is classically expensive. In practice, many chemically important multireference problems are framed through an active-space partitioning in which a subset of orbitals containing the dominant non-dynamical correlation is treated at high level, while the remaining orbitals are handled approximately. This logic fits naturally with quantum algorithms. A quantum processor need not describe the entire molecule at full many-body resolution to be chemically useful; instead, it may be deployed on the electronically active subspace where classical scaling is most severe.

Seen from this perspective, several recent VQE developments become especially meaningful. Adaptive ansatzes such as ADAPT-VQE reduce the variational space to operators that contribute most strongly to the current correlated state, which is potentially attractive when a compact but chemically expressive ansatz is needed for an active-space with substantial non-dynamical correlation. Orbital-optimized approaches such as ADAPT-VQE-SCF go one step further by allowing the orbital basis itself to respond to the multiconfigurational wavefunction. This is important because the compactness of a variational ansatz depends strongly on the orbital representation: a poorly chosen orbital basis can force the quantum circuit to encode correlation that could instead be absorbed into orbital rotations. Similarly, localized active-space and fragmentation-based methods such as LAS-nuVQE and FMO-VQE reflect the idea that many chemically relevant strongly correlated problems are spatially structured rather than uniformly delocalized. If correlation can be concentrated into chemically meaningful fragments or local active spaces, then quantum resources may be used selectively rather than globally.

This point is especially relevant for catalysis. Many catalytic mechanisms involve a localized reactive center embedded in a larger ligand environment or extended material. The chemically decisive physics is often concentrated in a modest subset of valence orbitals associated with bond activation, redox change, spin crossover, or metal-ligand reorganization, even when the full system is much larger. For transition-metal catalysis, examples include oxidative addition and reductive elimination, migratory insertion, proton-coupled electron transfer, spin-state changes, and the formation of highly active intermediates (e.g., oxo, nitrene, or carbene). These species frequently exhibit near-degenerate d-shell manifolds, open-shell configurations, and strong ligand-field sensitivity, making them difficult to describe with a single determinant and often expensive even for advanced classical multireference methods.

For such systems, the most plausible early role of VQE is not a full quantum treatment of an entire catalytic cycle, but a quantum-assisted active-space solver embedded within a broader classical workflow. In this setting, VQE would be used to capture the dominant non-dynamical correlation in the reactive subspace, while orbital optimization, embedding, fragmentation, and dynamical-correlation corrections remain classical. This division of labor is chemically sensible and mirrors the long-standing architecture of classical multireference methods, where the major challenge is often the combinatorial growth of the strongly correlated active-space wavefunction rather than the outer environment. If a quantum routine can improve the treatment of that subspace while preserving physically meaningful symmetries and maintaining manageable measurement cost, then it could become useful even before full end-to-end quantum advantage is realized.

The catalytic context also clarifies which VQE features matter most. First, orbital relaxation is likely essential. In transition-metal chemistry, the apparent multireference character depends sensitively on the orbital basis, and methods that combine VQE with orbital optimization may therefore be more attractive than fixed-orbital ansatzes. Second, localization and fragmentation are likely to be practical rather than optional. Realistic catalysts can be too large for monolithic quantum treatments in the near term, so methods that isolate reactive fragments or localized active spaces offer a clearer path to tractable calculations. Third, symmetry preservation matters chemically, not just formally. Spin, particle number, and point-group structure can strongly influence catalytic energetics and state ordering; ansatzes that violate these symmetries may become difficult to interpret even if they are circuit-efficient. Fourth, the key metric is not simply total energy error, but whether the method can deliver balanced relative energetics across multiple spin states and reaction intermediates and transition states.

This suggests a more concrete definition of chemically meaningful progress for VQE in catalysis. A compelling milestone would not be the reproduction of benchmark energies for small, single-reference molecules, but rather a calculation in which a quantum-assisted active-space treatment changes or verifies the mechanistic picture in a system where different classical approximations show inconsistency. Examples might include resolving a delicate spin-state ordering, distinguishing between competing pathways with different multiconfigurational character, or producing a more balanced description of bond activation in an open-shell transition-metal system. Even if such calculations remain embedded within classical orbital optimization, fragmentation, or perturbative correction frameworks, they would represent a more relevant demonstration of chemical value than isolated ground-state benchmarks on minimal-basis molecules.

At the same time, the obstacles should be stated plainly. Catalytic systems often combine exactly the features that stress VQE most severely: large active spaces, many relevant spin states, substantial orbital relaxation, and strong sensitivity to geometry and environment. Measurement overhead remains a central bottleneck, and the compactness gains of adaptive ansatzes can be offset by repeated gradient evaluation. Fragmentation and localized active-space approximations can reduce the quantum cost, but they introduce a second layer of chemical approximation whose validity depends on the degree of inter-fragment entanglement and electron delocalization. Moreover, the treatment of dynamical correlation remains a major open issue. Most chemistry-inspired VQE workflows are currently best viewed as methods for non-dynamical correlation, whereas catalytic accuracy often depends on recovering both non-dynamical and dynamical contributions in a balanced way. For this reason, some of the most important future developments may not be entirely new ansatzes, but rather improved hybrid schemes that combine VQE active-space solvers with perturbative, embedding, or coupled-cluster-like corrections.

Overall, we believe that multireference active-space chemistry is the clearest long-term scientific justification for VQE in molecular electronic structure, and catalysis provides some of its most compelling application targets. The most promising route is unlikely to be a standalone black-box VQE, but rather a chemistry-aware workflow in which quantum resources are focused on the strongly correlated subspace most responsible for mechanistic ambiguity.

\subsection{Excited states, photochemistry, and spectroscopy as a second major frontier}
Many of the molecular phenomena of greatest interest in spectroscopy, photochemistry, and photoactive materials depend not on a single electronic ground state, but on a manifold of low-lying excited states and their couplings. Absorption and emission processes, charge transfer, photo-isomerization, proton transfer, radiationless decay, and conical-intersection dynamics all require an electronic-structure description that remains balanced across multiple states, often over changing geometries. This makes excited-state chemistry a stringent test for variational quantum algorithms and, at the same time, a potentially important use case.

From a methodological standpoint, excited-state quantum chemistry is more demanding than ground-state energy minimization for at least three reasons. First, one usually needs more than one state, often several states of the same symmetry and spin multiplicity within a small energy window. Second, these states must be described consistently across nuclear geometries, especially when constructing potential energy surfaces or following nonadiabatic pathways. Third, chemically relevant observables may extend beyond state energies to include oscillator strengths, transition moments, rotational strengths, and other spectroscopic quantities. A method that obtains one excitation energy at a fixed geometry is therefore only a partial solution to the broader photochemical and spectroscopic problem.

These requirements help distinguish the excited-state VQE strategies discussed above. State-specific methods such as VQD and VQE/AC are conceptually straightforward and can be useful when only a few low-lying states are needed. Their main strength is direct targeting: each state is optimized variationally and can, in principle, be adapted carefully to the symmetry sector of interest. However, this sequential strategy can become cumbersome when many states are required, when surfaces must be followed over many geometries, or when errors in lower states propagate into higher ones. In a photochemical setting, where one may need smooth and balanced potential energy surfaces near crossings or conical intersections, such state-by-state optimization can become difficult to manage.

Subspace-based and response-based strategies are attractive precisely because they better reflect how excited states are used in chemistry. SS-VQE seeks a low-energy subspace rather than a single state, making it conceptually aligned with problems where several states of similar character matter simultaneously, as in absorption spectra or low-lying valence photochemistry. EOM-based methods go even further by importing the familiar linear-response logic of classical excited-state theory into the quantum-classical setting. Once a correlated ground-state reference has been prepared, a family of excitations can be obtained from measured reduced density matrices and a classical eigenvalue problem. This structure is appealing for spectroscopy because it offers a route to multiple excited states from a common reference, promoting internal balance among states without the need for repeated full variational optimizations. Orbital-optimized EOM variants are particularly interesting in this context because they combine active-space compression with access to response properties such as oscillator and rotational strengths, connecting more directly to experimentally measurable observables.

These distinctions matter strongly in photochemistry. In many photochemical processes, the key challenge is not the vertical excitation energy at a single equilibrium geometry, but the ability to describe changing state character and near-degeneracy along nuclear motion. Conical intersections, avoided crossings, excited-state proton transfer, and photo-isomerization all require a balanced treatment of multiple states whose ordering and orbital composition can change rapidly with geometry. This favors methods that are robust to state crossings and that preserve a chemically meaningful state manifold, rather than methods tuned only for isolated state energies. It also raises the importance of orbital optimization, spin symmetry, and state tracking. Excited-state methods that work well at equilibrium may become much less reliable if the underlying orbital basis is not suitable away from equilibrium or if the ansatz cannot represent multiple qualitatively distinct states with comparable fidelity.

For spectroscopy, the priorities are somewhat different but equally instructive. Here one often seeks not just excitation energies but families of transitions and associated intensities over a set of low-lying states (e.g., X-ray absorption spectroscopy). In this regime, subspace and EOM-like methods may ultimately be more attractive than repeated state-specific optimizations, especially if the goal is to compute absorption spectra, circular dichroism, or related optical observables in a consistent framework. The oo-VQE-qEOM direction is especially promising in this respect because it begins to connect VQE-based excited-state theory with spectroscopy simulations, where oscillator strengths, rotational strengths, and active-space orbital relaxation are central rather than optional. This link highlights the practical usefulness of an excited-state quantum algorithm to provide chemically interpretable observables and state manifolds relevant to experiment.

Potential application areas are broad. In photochemistry, promising targets include chromophores with strong multistate character, photoinduced proton-transfer systems, and small-to-moderate molecules exhibiting conical intersections or rapid geometry-driven state mixing. In spectroscopy, near-term interest may center on valence excitations in strongly correlated or multiconfigurational active spaces, where classical time-dependent density functional theory (TDDFT) may be qualitatively unreliable and where EOM-CC treatments become expensive or reference-dependent. In photocatalysis and photoactive materials, one can envision embedded or fragment-based quantum treatments of localized excitations, charge-transfer states, or open-shell excited intermediates, provided that the active subspace can be defined in a chemically sensible way.

As with ground-state multireference chemistry, however, realism requires caution. Excited-state VQE methods face several challenges beyond those of ground-state VQE. Orthogonality constraints, overlap estimation, and folded-spectrum constructions can introduce substantial sampling cost. EOM-based approaches reduce the need for repeated optimization but may require higher-order reduced density matrices or approximations whose impact must be assessed carefully. Smooth state tracking across geometries, which is essential for spectroscopy and photochemistry, remains underexplored relative to single-geometry benchmark calculations. Moreover, the combination of excited-state treatment with environmental effects, spin-orbit coupling, vibronic structure, and non-Born-Oppenheimer dynamics remains far from mature on quantum hardware. For near-term chemistry, the most realistic target is therefore not full photodynamics, but improved electronic-state treatment for selected low-lying manifolds in systems where classical methods already struggle to deliver balanced descriptions.

In our view, excited-state chemistry represents a second major frontier for chemically useful VQE because it naturally rewards exactly those features that hybrid quantum-classical methods may eventually provide: flexible treatment of strong correlation, explicit access to multistate manifolds, and integration with active-space or orbital-optimized workflows. The most promising early demonstrations are likely to be those that move beyond isolated excitation energies and instead show chemically meaningful improvement in state ordering, crossing behavior, or spectroscopic observables for problems with genuine multiconfigurational character.

\subsection{Realistic milestones for chemically useful VQE}
Taken together, the multireference and excited-state perspectives suggest that the most credible near-term role for VQE in chemistry is not universal replacement of classical electronic structure, but targeted deployment within workflows that isolate the electronically difficult part of the problem. For ground-state chemistry, this likely means quantum-assisted active-space treatments for open-shell, strongly correlated intermediates relevant to catalysis, coupled with classical embedding and dynamical-correlation corrections. For excited states, it likely means multistate treatments of low-lying manifolds in photochemical or spectroscopic applications where balanced state descriptions matter more than single-state absolute energies.

Accordingly, realistic milestones for the next stage of the field should be framed in chemical rather than purely algorithmic terms. For multireference chemistry, a meaningful milestone would be a VQE-based active-space calculation that resolves a chemically relevant ambiguity in a catalytic mechanism, spin-state ordering, or bond-activation problem beyond the reach of straightforward classical benchmarks. For excited states, a meaningful milestone would be a quantum-assisted treatment that improves the description of a low-lying state manifold, crossing region, or spectroscopic observable in a system where single-reference excited-state methods are unreliable. In both cases, success should be judged not only by absolute energy error, but by balanced relative energetics, robustness across geometries, symmetry fidelity, and compatibility with experimentally relevant observables.

This perspective also reinforces a broader methodological conclusion of this review: the most promising future for VQE in chemistry likely lies not in a single universally optimal ansatz, but in chemistry-aware hybrid workflows that combine quantum state preparation with orbital optimization, fragmentation, subspace construction, and classical post-processing. The systems most likely to justify these efforts are those where electronic complexity, rather than sheer molecular size alone, is the dominant barrier to classical computing.

\section*{Conflicts of interest}
There are no conflicts to declare.

\section*{Acknowledgements}
This work was supported by the U.S. Department of Energy (DOE), Office of Science (SC), Office of Basic Energy Sciences (BES), Division of Chemical Sciences, Geosciences, and Biosciences (CSGB) at Argonne National Laboratory under contract no. DE-AC02-06CH11357. T.H. and C.L. acknowledge the support by the DOE BES Clean Energy Technologies and Low-Carbon Manufacturing Initiative. R.K. and C.L. acknowledge the support by the DOE BES Computational Chemical Sciences Program. V.F.G. and C.L. acknowledge the support by the DOE BES CSGB Catalysis Science Program. The authors gratefully acknowledge the computing resources provided on Improv, a high-performance computing cluster operated by the Laboratory Computing Resource Center at Argonne National Laboratory.

\bibliography{bibliography}

@misc{kitaev1995,
      title={Quantum measurements and the Abelian Stabilizer Problem}, 
      author={A. Yu. Kitaev},
      year={1995},
      eprint={quant-ph/9511026},
      archivePrefix={arXiv},
      primaryClass={quant-ph},
      url={https://arxiv.org/abs/quant-ph/9511026}, 
}

@article{Abrams1999,
  author = {Abrams, D. S. and Lloyd, S.},
  title = {Quantum Algorithm Providing Exponential Speed Increase for Finding Eigenvalues and Eigenvectors},
  journal = {Physical Review Letters},
  volume = {83},
  number = {24},
  pages = {5162--5165},
  year = {1999},
  doi = {10.1103/PhysRevLett.83.5162}
}

@book{Nielsen2010,
  author = {Nielsen, M. A. and Chuang, I. L.},
  title = {Quantum Computation and Quantum Information},
  publisher = {Cambridge University Press},
  year = {2010}
}

@misc{Elfving2020,
      title={How will quantum computers provide an industrially relevant computational advantage in quantum chemistry?}, 
      author={V. E. Elfving and B. W. Broer and M. Webber and J. Gavartin and M. D. Halls and K. P. Lorton and A. Bochevarov},
      year={2020},
      eprint={2009.12472},
      archivePrefix={arXiv},
      primaryClass={quant-ph},
      url={https://arxiv.org/abs/2009.12472}, 
}

@article{Toyer2017,
 author = {Markus Reiher  and Nathan Wiebe  and Krysta M. Svore  and Dave Wecker  and Matthias Troyer },
title = {Elucidating reaction mechanisms on quantum computers},
journal = {Proceedings of the National Academy of Sciences},
volume = {114},
number = {29},
pages = {7555-7560},
year = {2017},
doi = {10.1073/pnas.1619152114},
URL = {https://www.pnas.org/doi/abs/10.1073/pnas.1619152114},
}

@misc{Aharonov1996,
title={Fault-Tolerant Quantum Computation With Constant Error Rate}, 
      author={Dorit Aharonov and Michael Ben-Or},
      year={1999},
      eprint={quant-ph/9906129},
      archivePrefix={arXiv},
      primaryClass={quant-ph},
      url={https://arxiv.org/abs/quant-ph/9906129}, 
}

@article{Dobsicek2007,
  author = {Dob{\v{s}}{\'i}{\v{c}}ek, M. and Johansson, G. and Shumeiko, V. and Wendin, G.},
  title = {Arbitrary Accuracy Iterative Quantum Phase Estimation Algorithm Using a Single Qubit},
  journal = {Physical Review A},
  volume = {76},
  number = {3},
  pages = {030306},
  year = {2007},
  doi = {10.1103/PhysRevA.76.030306}
}

@article{Wang2019,
  author = {Wang, D. and Higgott, O. and Brierley, S.},
  title = {Accelerated Variational Quantum Eigensolver},
  journal = {Physical Review Letters},
  volume = {122},
  number = {14},
  pages = {140504},
  year = {2019},
  doi = {10.1103/PhysRevLett.122.140504}
}

@article{Peruzzo2014,
  author = {Peruzzo, A. and McClean, J. and Shadbolt, P. and Yung, M.-H. and Zhou, X.-Q. and Love, P. J. and Aspuru-Guzik, A. and O’Brien, J. L.},
  title = {A Variational Eigenvalue Solver on a Photonic Quantum Processor},
  journal = {Nature Communications},
  volume = {5},
  number = {1},
  pages = {4213},
  year = {2014},
  doi = {10.1038/ncomms5213}
}

@article{McClean_2014,
   title={Exploiting Locality in Quantum Computation for Quantum Chemistry},
   volume={5},
   ISSN={1948-7185},
   url={http://dx.doi.org/10.1021/jz501649m},
   DOI={10.1021/jz501649m},
   number={24},
   journal={The Journal of Physical Chemistry Letters},
   publisher={American Chemical Society (ACS)},
   author={McClean, Jarrod R. and Babbush, Ryan and Love, Peter J. and Aspuru-Guzik, Alán},
   year={2014},
   month=dec, pages={4368–4380} }

@article{Grimsley2019,
  author = {Grimsley, H. R. and Economou, S. E. and Barnes, E. and Mayhall, N. J.},
  title = {An Adaptive Variational Algorithm for Exact Molecular Simulations on a Quantum Computer},
  journal = {Nature Communications},
  volume = {10},
  number = {1},
  pages = {3007},
  year = {2019},
  doi = {10.1038/s41467-019-10988-2}
}

@article{Preskill2018,
  author = {Preskill, J.},
  title = {Quantum Computing in the NISQ Era and Beyond},
  journal = {Quantum},
  volume = {2},
  pages = {79},
  year = {2018},
  doi = {10.22331/q-2018-08-06-79}
}

@article{Tang2021,
  author = {Tang, H. L. and Shkolnikov, V. O. and Barron, G. S. and Grimsley, H. R. and Mayhall, N. J. and Barnes, E. and Economou, S. E.},
  title = {Qubit-ADAPT-VQE: An Adaptive Algorithm for Constructing Hardware-Efficient Ansätze on a Quantum Processor},
  journal = {PRX Quantum},
  volume = {2},
  number = {2},
  pages = {020310},
  year = {2021},
  doi = {10.1103/PRXQuantum.2.020310}
}

@article{Fitzpatrick_2024,
   title={Self-Consistent Field Approach for the Variational Quantum Eigensolver: Orbital Optimization Goes Adaptive},
   volume={128},
   ISSN={1520-5215},
   url={http://dx.doi.org/10.1021/acs.jpca.3c05882},
   DOI={10.1021/acs.jpca.3c05882},
   number={14},
   journal={The Journal of Physical Chemistry A},
   publisher={American Chemical Society (ACS)},
   author={Fitzpatrick, Aaron and Nykänen, Anton and Talarico, N. Walter and Lunghi, Alessandro and Maniscalco, Sabrina and García-Pérez, Guillermo and Knecht, Stefan},
   year={2024},
   month=mar, pages={2843–2856} }

@article{Yordanov2021,
  author = {Yordanov, Y. S. and Arvidsson-Shukur, D. R. M. and Barnes, C. H. W.},
  title = {Efficient Quantum Circuits for Quantum Computational Chemistry},
  journal = {Physical Review A},
  volume = {103},
  number = {3},
  pages = {032610},
  year = {2021},
  doi = {10.1103/PhysRevA.103.032610}
}

@article{Benfenati2021,
  author = {Benfenati, Francesco and Mazzola, Guglielmo and Capecci, Chiara and Barkoutsos, Panagiotis Kl. and Ollitrault, Pauline J. and Tavernelli, Ivano and Guidoni, Leonardo},
  title = {Improved Accuracy on Noisy Devices by Non-Unitary Variational Quantum Eigensolver for Chemistry Applications},
  journal = {arXiv preprint arXiv:2101.09316},
  year = {2021},
  pages = {1},
  eprint = {2101.09316},
  eprinttype = {arXiv}
}

@article{Tilly2022,
  author = {Tilly, J. and Chen, H. and Cao, S. and Picozzi, D. and Setia, K. and Li, Y. and Grant, E. and Wossnig, L. and Rungger, I. and Booth, G. H. and Tennyson, J.},
  title = {The Variational Quantum Eigensolver: A Review of Methods and Best Practices},
  journal = {Physics Reports},
  volume = {986},
  pages = {1--128},
  year = {2022},
  doi = {10.1016/j.physrep.2022.08.003}
}

@article{wang2025,
  title={Nonunitary Variational Quantum Eigensolver with the Localized Active Space Method and Cost Mitigation},
  author={Wang, Qiaohong and D’Cunha, Ruhee and Mitra, Abhishek and Gray, Stephen K and Otten, Matthew and Gagliardi, Laura},
  journal={The Journal of Physical Chemistry A},
  volume={129},
  number={34},
  pages={7999--8012},
  year={2025},
  publisher={ACS Publications}
}

@article{Fedorov2022,
  author = {Fedorov, Dmitry A. and Alexeev, Yuri and Gray, Stephen K. and Otten, Matthew J.},
  title = {Unitary Selective Coupled-Cluster Method},
  journal = {Quantum},
  volume = {6},
  pages = {703},
  year = {2022},
  doi = {10.22331/q-2022-04-18-703},
  eprint = {2109.12652},
  eprinttype = {arXiv}
}

@article{Holmes2016,
  author = {Holmes, Adam A. and Tubman, Norm M. and Umrigar, C. J.},
  title = {Heat-Bath Configuration Interaction: An Efficient Selected Configuration Interaction Algorithm Inspired by Heat-Bath Sampling},
  journal = {Journal of Chemical Theory and Computation},
  year = {2016},
  volume = {12},
  number = {8},
  pages = {3674--3680},
  doi = {10.1021/acs.jctc.6b00407}
}

@article{Zhang2022,
  author = {Zhang, Yu and Cincio, Lukasz and Negre, Christian F. A. and Czarnik, Piotr and Coles, Patrick J. and Anisimov, Petr M. and Mniszewski, Susan M. and Tretiak, Sergei and Dub, Pavel A.},
  title = {Variational Quantum Eigensolver with Reduced Circuit Complexity},
  journal = {npj Quantum Information},
  volume = {8},
  number = {1},
  pages = {96},
  year = {2022},
  doi = {10.1038/s41534-022-00599-z}
}

@article{Rissler_2006,
   title={Measuring orbital interaction using quantum information theory},
   volume={323},
   ISSN={0301-0104},
   url={http://dx.doi.org/10.1016/j.chemphys.2005.10.018},
   DOI={10.1016/j.chemphys.2005.10.018},
   number={2–3},
   journal={Chemical Physics},
   publisher={Elsevier BV},
   author={Rissler, Jörg and Noack, Reinhard M. and White, Steven R.},
   year={2006},
   month=apr, pages={519–531} }

@article{Barca2020,
  author = {Barca, G. M. J. and Bertoni, C. and Carrington, L. and Datta, D. and De Silva, N. and Deustua, J. E. and Fedorov, D. G. and Gour, J. R. and Gunina, A. O. and Guidez, E. and Harville, T. and Irle, S. and Ivanic, J. and Kowalski, K. and Leang, S. S. and Li, H. and Li, W. and Lutz, J. J. and Magoulas, I. and Mato, J. and Mironov, V. and Nakata, H. and Pham, B. Q. and Piecuch, P. and Poole, D. and Pruitt, S. R. and Rendell, A. P. and Roskop, L. B. and Ruedenberg, K. and Sattasathuchana, T. and Schmidt, M. W. and Shen, J. and Slipchenko, L. and Sosonkina, M. and Sundriyal, V. and Tiwari, A. and Vallejo, J. L. G. and Westheimer, B. and Włoch, M. and Xu, P. and Zahariev, F. and Gordon, M. S.},
  title = {Recent developments in the general atomic and molecular electronic structure system},
  journal = {J. Chem. Phys.},
  volume = {152},
  number = {15},
  pages = {154102},
  year = {2020},
  doi = {10.1063/5.0005188}
}

@article{Lim2024,
  author = {Lim, Hocheol and Kang, Doo Hyung and Kim, Jeonghoon and Pellow-Jarman, Aidan and McFarthing, Shane and Pellow-Jarman, Rowan and Jeon, Hyeon-Nae and Oh, Byungdu and Oh, Shane and Rhee, June-Koo Kevin and No, Kyoung Tai},
  title = {Fragment Molecular Orbital-Based Variational Quantum Eigensolver for Quantum Chemistry in the Age of Quantum Computing},
  journal = {Scientific Reports},
  volume = {14},
  number = {1},
  pages = {2422},
  year = {2024},
  doi = {10.1038/s41598-024-52926-3}
}

@article{Nakano2000,
  author = {Nakano, T. and Kaminuma, T. and Sato, T. and Akiyama, Y. and Uebayasi, M. and Kitaura, K.},
  title = {Fragment molecular orbital method: application to molecular dynamics simulation of proteins},
  journal = {Chem. Phys. Lett.},
  volume = {318},
  number = {6},
  pages = {614--618},
  year = {2000},
  doi = {10.1016/S0009-2614(00)00093-7}
}

@article{Roos1980,
  author = {Roos, Bj{\"o}rn O. and Taylor, Peter R. and Sigbahn, Per E. M.},
  title = {A Complete Active Space SCF Method (CASSCF) Using a Density Matrix Formulated Super-CI Approach},
  journal = {Chemical Physics},
  volume = {48},
  number = {2},
  pages = {157--173},
  year = {1980},
  doi = {10.1016/0301-0104(80)80045-0}
}

@article{Siegbahn1981,
  author = {Siegbahn, Per E. M. and Heiberg, Astrid and Roos, Bj{\"o}rn O. and Levy, Bernard},
  title = {A Comparison of the Super-CI and the Newton-Raphson Scheme in the Complete Active Space SCF Method},
  journal = {Physica Scripta},
  volume = {21},
  number = {3-4},
  pages = {323},
  year = {1981},
  doi = {10.1088/0031-8949/21/3-4/014}
}

@article{Higgott2019variationalquantum,
  doi = {10.22331/q-2019-07-01-156},
  url = {https://doi.org/10.22331/q-2019-07-01-156},
  title = {Variational {Q}uantum {C}omputation of {E}xcited {S}tates},
  author = {Higgott, Oscar and Wang, Daochen and Brierley, Stephen},
  journal = {{Quantum}},
  issn = {2521-327X},
  publisher = {{Verein zur F{\"{o}}rderung des Open Access Publizierens in den Quantenwissenschaften}},
  volume = {3},
  pages = {156},
  month = jul,
  year = {2019}
}

@article{cadi2024folded,
  title={Folded spectrum vqe: A quantum computing method for the calculation of molecular excited states},
  author={Cadi Tazi, Lila and Thom, Alex JW},
  journal={Journal of Chemical Theory and Computation},
  volume={20},
  number={6},
  pages={2491--2504},
  year={2024},
  publisher={ACS Publications}
}

@article{ollitrault2020quantum,
  title={Quantum equation of motion for computing molecular excitation energies on a noisy quantum processor},
  author={Ollitrault, Pauline J and Kandala, Abhinav and Chen, Chun-Fu and Barkoutsos, Panagiotis Kl and Mezzacapo, Antonio and Pistoia, Marco and Sheldon, Sarah and Woerner, Stefan and Gambetta, Jay M and Tavernelli, Ivano},
  journal={Physical Review Research},
  volume={2},
  number={4},
  pages={043140},
  year={2020},
  publisher={APS}
}

@article{pavosevic2021multicomponent,
  title={Multicomponent unitary coupled cluster and equation-of-motion for quantum computation},
  author={Pavosevic, Fabijan and Hammes-Schiffer, Sharon},
  journal={Journal of Chemical Theory and Computation},
  volume={17},
  number={6},
  pages={3252--3258},
  year={2021},
  publisher={ACS Publications}
}

@article{gocho2023excited,
  title={Excited state calculations using variational quantum eigensolver with spin-restricted ans{\"a}tze and automatically-adjusted constraints},
  author={Gocho, Shigeki and Nakamura, Hajime and Kanno, Shu and Gao, Qi and Kobayashi, Takao and Inagaki, Taichi and Hatanaka, Miho},
  journal={npj Computational Materials},
  volume={9},
  number={1},
  pages={13},
  year={2023},
  publisher={Nature Publishing Group UK London}
}

@article{asthana2023quantum,
  title={Quantum self-consistent equation-of-motion method for computing molecular excitation energies, ionization potentials, and electron affinities on a quantum computer},
  author={Asthana, Ayush and Kumar, Ashutosh and Abraham, Vibin and Grimsley, Harper and Zhang, Yu and Cincio, Lukasz and Tretiak, Sergei and Dub, Pavel A and Economou, Sophia E and Barnes, Edwin and others},
  journal={Chemical Science},
  volume={14},
  number={9},
  pages={2405--2418},
  year={2023},
  publisher={Royal Society of Chemistry}
}

@article{jensen2024quantum,
  title={Quantum equation of motion with orbital optimization for computing molecular properties in near-term quantum computing},
  author={Jensen, Phillip WK and Kjellgren, Erik Rosendahl and Reinholdt, Peter and Ziems, Karl Michael and Coriani, Sonia and Kongsted, Jacob and Sauer, Stephan PA},
  journal={Journal of Chemical Theory and Computation},
  volume={20},
  number={9},
  pages={3613--3625},
  year={2024},
  publisher={ACS Publications}
}

@article{Feynman1982,
  author = {Feynman, Richard P.},
  title = {Simulating physics with computers},
  journal = {International Journal of Theoretical Physics},
  volume = {21},
  number = {6-7},
  pages = {467--488},
  year = {1982},
  doi = {10.1007/BF02650179}
}

@article{Feniou_2023,
   title={Overlap-ADAPT-VQE: practical quantum chemistry on quantum computers via overlap-guided compact Ansätze},
   volume={6},
   ISSN={2399-3650},
   url={http://dx.doi.org/10.1038/s42005-023-01312-y},
   DOI={10.1038/s42005-023-01312-y},
   pages ={1},
   journal={Communications Physics},
   publisher={Springer Science and Business Media LLC},
   author={Feniou, César and Hassan, Muhammad and Traoré, Diata and Giner, Emmanuel and Maday, Yvon and Piquemal, Jean-Philip},
   year={2023},
   month=jul }

@article{Feniou2025,
author = {Feniou, C{'e}sar and Hassan, Muhammad and Claudon, Baptiste and Courtat, Axel and Adjoua, Olivier and Maday, Yvon and Piquemal, Jean-Philip},
title = {Greedy gradient-free adaptive variational quantum algorithms on a noisy intermediate scale quantum computer},
journal = {Scientific Reports},
pages ={18689},
volume = {15},
year = {2025},
doi = {10.1038/s41598-025-99962-1},
publisher = {Springer Nature}
}

@article{lan2022amplitude,
  author = {Lan, Zhihao and Liang, WanZhen},
  title = {Amplitude Reordering Accelerates the Adaptive Variational Quantum Eigensolver Algorithms},
  journal = {Journal of Chemical Theory and Computation},
  volume = {18},
  number = {9},
  pages = {5267--5275},
  year = {2022},
  doi = {10.1021/acs.jctc.2c00403},
  url = {https://doi.org/10.1021/acs.jctc.2c00403},
  publisher = {American Chemical Society}
}

@misc{zhang2025diffusionenhancedoptimizationvariationalquantum,
      title={Diffusion-Enhanced Optimization of Variational Quantum Eigensolver for General Hamiltonians}, 
      author={Shikun Zhang and Zheng Qin and Yongyou Zhang and Yang Zhou and Rui Li and Chunxiao Du and Zhisong Xiao},
      year={2025},
      eprint={2501.05666},
      archivePrefix={arXiv},
      primaryClass={quant-ph},
      url={https://arxiv.org/abs/2501.05666}, 
}

@misc{shi2025efficienthamiltonianawarequantumnatural,
      title={Efficient Hamiltonian-aware Quantum Natural Gradient Descent for Variational Quantum Eigensolvers}, 
      author={Chenyu Shi and Hao Wang},
      year={2025},
      eprint={2511.14511},
      archivePrefix={arXiv},
      primaryClass={quant-ph},
      url={https://arxiv.org/abs/2511.14511}, 
}

@article{iQCC,
author = {Ryabinkin, Ilya G. and Lang, Robert A. and Genin, Scott N. and Izmaylov, Artur F.},
title = {Iterative Qubit Coupled Cluster Approach with Efficient Screening of Generators},
journal = {Journal of Chemical Theory and Computation},
volume = {16},
number = {2},
pages = {1055-1063},
year = {2020},
doi = {10.1021/acs.jctc.9b01084},
}

@article{Levine2020,
author = {Levine, Daniel S. and Hait, Diptarka and Tubman, Norm M. and Lehtola, Susi and Whaley, K. Birgitta and Head-Gordon, Martin},
title = {CASSCF with Extremely Large Active Spaces Using the Adaptive Sampling Configuration Interaction Method},
journal = {Journal of Chemical Theory and Computation},
volume = {16},
number = {4},
pages = {2340-2354},
year = {2020},
doi = {10.1021/acs.jctc.9b01255},
    note ={PMID: 32109055},
URL = {https://doi.org/10.1021/acs.jctc.9b01255},
eprint = {https://doi.org/10.1021/acs.jctc.9b01255}
}

@article{Montgomery2018,
author = {Montgomery, Jason
M. and Mazziotti, David A.},
title = {Strong Electron Correlation in Nitrogenase Cofactor, FeMoco},
journal = {The Journal of Physical Chemistry A},
volume = {122},
number = {22},
pages = {4988-4996},
year = {2018},
doi = {10.1021/acs.jpca.8b00941},
    note ={PMID: 29771514},
URL = {https://doi.org/10.1021/acs.jpca.8b00941},
eprint = {https://doi.org/10.1021/acs.jpca.8b00941}
}

@article{Markus2017,
author = {Markus Reiher  and Nathan Wiebe  and Krysta M. Svore  and Dave Wecker  and Matthias Troyer },
title = {Elucidating reaction mechanisms on quantum computers},
journal = {Proceedings of the National Academy of Sciences},
volume = {114},
number = {29},
pages = {7555-7560},
year = {2017},
doi = {10.1073/pnas.1619152114},
URL = {https://www.pnas.org/doi/abs/10.1073/pnas.1619152114},
eprint = {https://www.pnas.org/doi/pdf/10.1073/pnas.1619152114},
}

@article{Bal2024,
  author  = {Bal, Mustafa and Murthy, Akshay A. and Zhu, Shaojiang and Crisa, Francesco and others},
  title   = {Systematic improvements in transmon qubit coherence enabled by niobium surface encapsulation},
  journal = {npj Quantum Information},
  volume  = {10},
  pages   = {43},
  year    = {2024},
  doi     = {10.1038/s41534-024-00840-x},
  url     = {https://doi.org/10.1038/s41534-024-00840-x}
}

@article{Jiang2025,
    author = {Jiang, Yao-Yao and Deng, Chunqing and Fan, Heng and Li, Bing-Yang and Sun, Luyan and Tan, Xin-Sheng and Wang, Weiting and Xue, Guang-Ming and Yan, Fei and Yu, Hai-Feng and Zhang, Ying-Shan and Zhang, Yu-Ran and Zou, Chang-Ling},
    title = {Advancements in superconducting quantum computing},
    journal = {National Science Review},
    volume = {12},
    number = {8},
    pages = {nwaf246},
    year = {2025},
    month = {06},
    issn = {2095-5138},
    doi = {10.1093/nsr/nwaf246},
    url = {https://doi.org/10.1093/nsr/nwaf246},
    eprint = {https://academic.oup.com/nsr/article-pdf/12/8/nwaf246/63509341/nwaf246.pdf},
}

@article{
Awshalom2025,
author = {David D. Awschalom  and Hannes Bernien  and Ronald Hanson  and William D. Oliver  and Jelena Vučković },
title = {Challenges and opportunities for quantum information hardware},
journal = {Science},
volume = {390},
number = {6777},
pages = {1004-1010},
year = {2025},
doi = {10.1126/science.adz8659},
URL = {https://www.science.org/doi/abs/10.1126/science.adz8659},
eprint = {https://www.science.org/doi/pdf/10.1126/science.adz8659},
}

@article{Ramoa2025,
  author = {Ramôa, M. and others},
  title = {Reducing the resources required by ADAPT-VQE using coupled exchange operators and improved subroutines},
  journal = {npj Quantum Inf.},
  volume = {11},
  pages = {86},
  year = {2025},
  doi = {10.1038/s41534-025-01039-4}
}

@article{Federov2008,
author = {Fedorov, Dmitri G. and Jensen, Jan H. and Deka, Ramesh C. and Kitaura, Kazuo},
title = {Covalent Bond Fragmentation Suitable To Describe Solids in the Fragment Molecular Orbital Method},
journal = {The Journal of Physical Chemistry A},
volume = {112},
number = {46},
pages = {11808-11816},
year = {2008},
doi = {10.1021/jp805435n},
    note ={PMID: 18942816},
URL = {https://doi.org/10.1021/jp805435n},
eprint = {https://doi.org/10.1021/jp805435n}
}

@article{Bauer2020,
author = {Bauer, Bela and Bravyi, Sergey and Motta, Mario and Chan, Garnet Kin-Lic},
title = {Quantum Algorithms for Quantum Chemistry and Quantum Materials Science},
journal = {Chemical Reviews},
volume = {120},
number = {22},
pages = {12685-12717},
year = {2020},
doi = {10.1021/acs.chemrev.9b00829},
    note ={PMID: 33090772},
URL = {https://doi.org/10.1021/acs.chemrev.9b00829},
eprint = {https://doi.org/10.1021/acs.chemrev.9b00829}
}

@article{Rishu_1,
author = {Khurana, Rishu and Liu, Cong},
title = {Unveiling the Redox Noninnocence of Metallocorroles: Exploring K-Edge X-ray Absorption Near-Edge Spectroscopy with a Multiconfigurational Wave Function Approach},
journal = {The Journal of Physical Chemistry Letters},
volume = {15},
number = {44},
pages = {10985-10995},
year = {2024},
doi = {10.1021/acs.jpclett.4c02410},
URL = {https://doi.org/10.1021/acs.jpclett.4c02410},
}

@article{Potter2015,
author = {Potter, Matthew E. and Paterson, A. James and Mishra, Bhoopesh and Kelly, Shelly D. and Bare, Simon R. and Corà, Furio and Levy, Alan B. and Raja, Robert},
title = {Spectroscopic and Computational Insights on Catalytic Synergy in Bimetallic Aluminophosphate Catalysts},
journal = {Journal of the American Chemical Society},
volume = {137},
number = {26},
pages = {8534-8540},
year = {2015},
doi = {10.1021/jacs.5b03734},
URL = {https://doi.org/10.1021/jacs.5b03734},
}

@article{Patel2022,
author = {Patel, Prajay and Lu, Zheng and Jafari, Mehrafshan G. and Hernández-Prieto, Cristina and Zatsepin, Pavel and Mindiola, Daniel J. and Kaphan, David M. and Delferro, Massimiliano and Kropf, A. Jeremy and Liu, Cong},
title = {Integrated Experimental and Computational K-Edge X-ray Absorption Near-Edge Structure Analysis of Vanadium Catalysts},
journal = {The Journal of Physical Chemistry C},
volume = {126},
number = {29},
pages = {11949-11962},
year = {2022},
doi = {10.1021/acs.jpcc.2c02049},
URL = {https://doi.org/10.1021/acs.jpcc.2c02049},
}

@article{Rishu_2025,
AUTHOR = {Khurana, Rishu and Liu, Cong},
TITLE = {M-Edge Spectroscopy of Transition Metals: Principles, Advances, and Applications},
JOURNAL = {Catalysts},
VOLUME = {15},
year = {2025},
number = {8},
ARTICLE-NUMBER = {722},
pages = {1-13},
URL = {https://www.mdpi.com/2073-4344/15/8/722},
DOI = {10.3390/catal15080722}
}

@article{Rishu_2024,
author = {Khurana, Rishu and Agarawal, Valay and Liu, Cong},
title = {Exploring the Computational Aspects of Propylene Oligomerization Catalysis Using M2M Type Trimetallic MOF Nodes},
journal = {The Journal of Physical Chemistry C},
volume = {128},
number = {40},
pages = {16986-16995},
year = {2024},
doi = {10.1021/acs.jpcc.4c04992},
URL = { https://doi.org/10.1021/acs.jpcc.4c04992},
}

@article{Gagliardi_2019,
author = {Gaggioli, Carlo
Alberto and Stoneburner, Samuel J. and Cramer, Christopher J. and Gagliardi, Laura},
title = {Beyond Density Functional Theory: The Multiconfigurational Approach To Model Heterogeneous Catalysis},
journal = {ACS Catalysis},
volume = {9},
number = {9},
pages = {8481-8502},
year = {2019},
doi = {10.1021/acscatal.9b01775},

URL = {https://doi.org/10.1021/acscatal.9b01775},
}

@article{Truhlar_2022,
    author = {Zhou, Chen and Hermes, Matthew R. and Wu, Dihua and Bao, Jie J. and Pandharkar, Riddhish and King, Daniel S. and Zhang, Dayou and Scott, Thais R. and Lykhin, Aleksandr O. and Gagliardi, Laura and Truhlar, Donald G.},
    title = {Electronic structure of strongly correlated systems: recent developments in multiconfiguration pair-density functional theory and multiconfiguration nonclassical-energy functional theory},
    journal = {Chemical Science},
    volume = {13},
    number = {26},
    pages = {7685-7706},
    year = {2022},
    month = {07},
    doi = {10.1039/d2sc01022d},
    url = {https://doi.org/10.1039/d2sc01022d},
}

@article{Gagliardi_2025,
author = {Mandal, Mukunda and Hermes, Matthew R. and Berger, Fabian and Sauer, Joachim and Gagliardi, Laura},
title = {Modeling Oxidative Dehydrogenation of Propane with Supported Vanadia Catalysts Using Multireference Methods},
journal = {The Journal of Physical Chemistry C},
volume = {129},
number = {32},
pages = {14418-14429},
year = {2025},
doi = {10.1021/acs.jpcc.5c04695},

URL = {https://doi.org/10.1021/acs.jpcc.5c04695},
}

@article{Gagliardi_2022,
author = {Stroscio, Gautam D. and Zhou, Chen and Truhlar, Donald G. and Gagliardi, Laura},
title = {Multiconfiguration Pair-Density Functional Theory Calculations of Iron(II) Porphyrin: Effects of Hybrid Pair-Density Functionals and Expanded RAS and DMRG Active Spaces on Spin-State Orderings},
journal = {The Journal of Physical Chemistry A},
volume = {126},
number = {24},
pages = {3957-3963},
year = {2022},
doi = {10.1021/acs.jpca.2c02347},
    note ={PMID: 35674705},

URL = {https://doi.org/10.1021/acs.jpca.2c02347},
}

@article{nakanishi2019subspace,
  title={Subspace-search variational quantum eigensolver for excited states},
  author={Nakanishi, Ken M and Mitarai, Kosuke and Fujii, Keisuke},
  journal={Physical Review Research},
  volume={1},
  number={3},
  pages={033062},
  year={2019},
  publisher={APS}
}

@article{khurana2026multireference,
  author       = {Khurana, Rishu and Hermes, Matthew and Agarawal, Valay and Knight, Christopher and Gagliardi, Laura and Liu, Cong},
  title        = {Multireference Investigations of Ethylene Hydrogenation over Bimetallic Catalysts},
  year         = {2026},
  month        = may,
  day          = {27},
  note         = {Preprint, Version 1, available at Research Square},
  journal    = {Preprint, Research Square},
  doi          = {10.21203/rs.3.rs-9684388/v1},
  url          = {https://doi.org/10.21203/rs.3.rs-9684388/v1}
}

@article{Gagliardi_2026,
author = {Wardzala, Jacob J. and Kim, Younghwan and Khurana, Rishu and Mandal, Mukunda and Delferro, Massimiliano and Liu, Cong and Gagliardi, Laura},
title = {From Oxo to Oxyl to Biradical: Systematic Multireference Calculations of Methane Activation at MOF Nodes},
journal = {Journal of the American Chemical Society},
volume = {0},
number = {0},
pages = {null},
year = {0},
doi = {10.1021/jacs.6c05408},
note ={PMID: 42345126},

URL = {https://doi.org/10.1021/jacs.6c05408}
}

\end{document}